\documentclass[12pt,preprint]{aastex631}

\usepackage{natbib}
\usepackage{xcolor}
 
\newcommand\species[2]{#1 {\sc #2}}

\def\ie{\mbox{i.e.}}
\def\eg{\mbox{e.g.}}

\def\teff{\mbox{$T_{\rm eff}$}}
\def\logg{\mbox{log~{\it g}}}
\def\kmsec{\mbox{km~s$^{\rm -1}$}}

\def\vmicro{$V_{mic}$}
\def\sigmax{$\sigma_{Max}$}
\shorttitle{RRab Stars Rising Light}
\shortauthors{Preston et al.}

\begin{document}

\title{HYDROGEN AND HELIUM SHOCK PHENOMENA DURING RISING LIGHT 
       IN RR LYRAE FUNDAMENTAL MODE PULSATORS}

\correspondingauthor{Christopher Sneden}
\email{chris@verdi.as.utexas.edu}

\author{George W. Preston}
\affiliation{Carnegie Observatories, 813 Santa Barbara Street, Pasadena,
             CA 91101, USA; gwp,ian,shec@obs.carnegiescience.edu}
\author{Christopher Sneden}
\affiliation{Department of Astronomy and McDonald Observatory,
             The University of Texas, Austin, TX 78712, USA;
             chris@verdi.as.utexas.edu}
\author{Merieme Chadid}
\affiliation{University of C\^ote d’Azur, Nice Sophia--Antipolis University, 
             CNRS--UMR 7250, CS 34229, 06304 NICE Cedex 4, France; 
             chadid@unice.fr}

\begin{abstract}

We present measurements of H and He emission and absorption 
lines produced in RRab fundamental mode pulsators during primary light rise.
The lines define universal progressions of rise and decay in metal-poor 
RRab stars. 
Such a progression cannot be constructed for He in metal-rich RRab (those 
with [Fe/H]~$>$ $-$0.8) because weak \species{He}{i} emission is detected 
in only two of the six metal-rich RRab in our survey.
Great variety exists in the phase variations of the blue- and red-shifted 
absorption components of the 5876~\AA\ line during pre- and post-emission 
phases. 
Detection of measurable \species{He}{ii} 4686~\AA\ emission in eight RRab, 
three of them Blazhko variables, provides an additional constraint on 
ionization of Helium.

\end{abstract}

\section{INTRODUCTION}\label{intro}

Two weakly-visible H$\alpha$ emission episodes attributed to shock phenomena 
occur in RRab stars during declining light.   
The first, dubbed the ``third apparition'' \citep{preston11}, is identified 
by a weak red-shifted H$\alpha$ emission wing visible during about one third 
of the pulsation cycle following maximum light.  
Ad hoc explanations of this apparition have been advanced by \cite{chadid13}
and  \cite{gillet17,gillet19}.  
A second shock, identified by violet-shifted H$\alpha$ emission during the 
``bump'' preceding minimum light \citep{gillet88}, is discussed at length in 
the foregoing references. 
The existence of this shock was anticipated theoretically by the calculations 
of \cite{hill72}.  
Both emissions are accompanied by absorption line-doubling produced in 
outflowing and infalling atmospheric layers. 

\cite{chadid14} detected multi-shocks propagating through the atmosphere of 
RR\,Lyrae stars.
They identified new light curve properties\footnote{
nicknamed $jump$, $lump$, $rump$, $bump$, and $hump$, as introduced by 
\citealt{christy66} and expanded by \citealt{chadid14}}
induced by five shock waves, with different amplitudes and origins.
Recently \cite{prudil20} have used broad band photometry to identify (hump)
shock phenomena in a large sample of field RRab.
Metallic-line absorption phenomena, particularly shock-induced increases in 
macroturbulence during rising in stable RRab, were reviewed by 
\cite{preston09}, and studied in great detail more recently by \cite{preston19}.

This paper deals with strong shock phenomena (``first apparition'',
the hump) during
 primary light-rise, discovered 
spectroscopically by \cite{sanford49} and interpreted soon thereafter 
by \cite{schwarzschild52}.
All previous discussions of this primary shock in RR Lyrae stars have been 
based solely on the behaviors of Balmer emission and absorption lines
in the context of the \citeauthor{schwarzschild52} model.  
\cite{wallerstein59b} explored shock conditions in the Pop~II 
Cepheid W~Virginis.  
He calculated, by taking into account the energy of ionization in application 
of the Rankine-Hugoniot relations, that a shock velocity higher than 80~\kmsec\ 
is able to ionize Hydrogen and neutral Helium and, additionally, a small 
fraction of singly-ionized Helium. 
This condition is fulfilled by typical RRab stars (to be discussed in 
\S\ref{helium1}), 
so it came as no great surprise that 
\cite{preston09,preston11} detected \species{He}{i} emission and absorption 
lines during light-rise in 10 RRab stars and very weak \species{He}{ii} 
4685.68~\AA\ emission in 3 RRab stars.  
These H and He emissions, produced in the wakes of shock waves, can provide 
quantitative insight into the stratification of the outflowing and 
infalling atmospheric layers. 

In this paper we present new information about the time evolution of the 
pulsation velocities, equivalent widths, and line-of-sight broadening of the 
emission and absorption of the most prominent lines of H, \species{He}{i} and 
\species{He}{ii} in our spectra.
The format of the paper is as follows.  
In \S\ref{spectra} we describe the data set; in \S\ref{measures} we present 
Hydrogen and Helium emission and absorption line measurements; and in
\S\ref{hepersistence} we discuss issues raised by the behavior of the
\species{He}{i} $\lambda$5876 line.
\S\ref{conclusions} contains our concluding remarks.

\vspace*{0.2in}
\section{THE SPECTROSCOPIC DATA SET}\label{spectra}

\begin{figure}
\begin{center}
\includegraphics[scale=0.65,angle=-90]{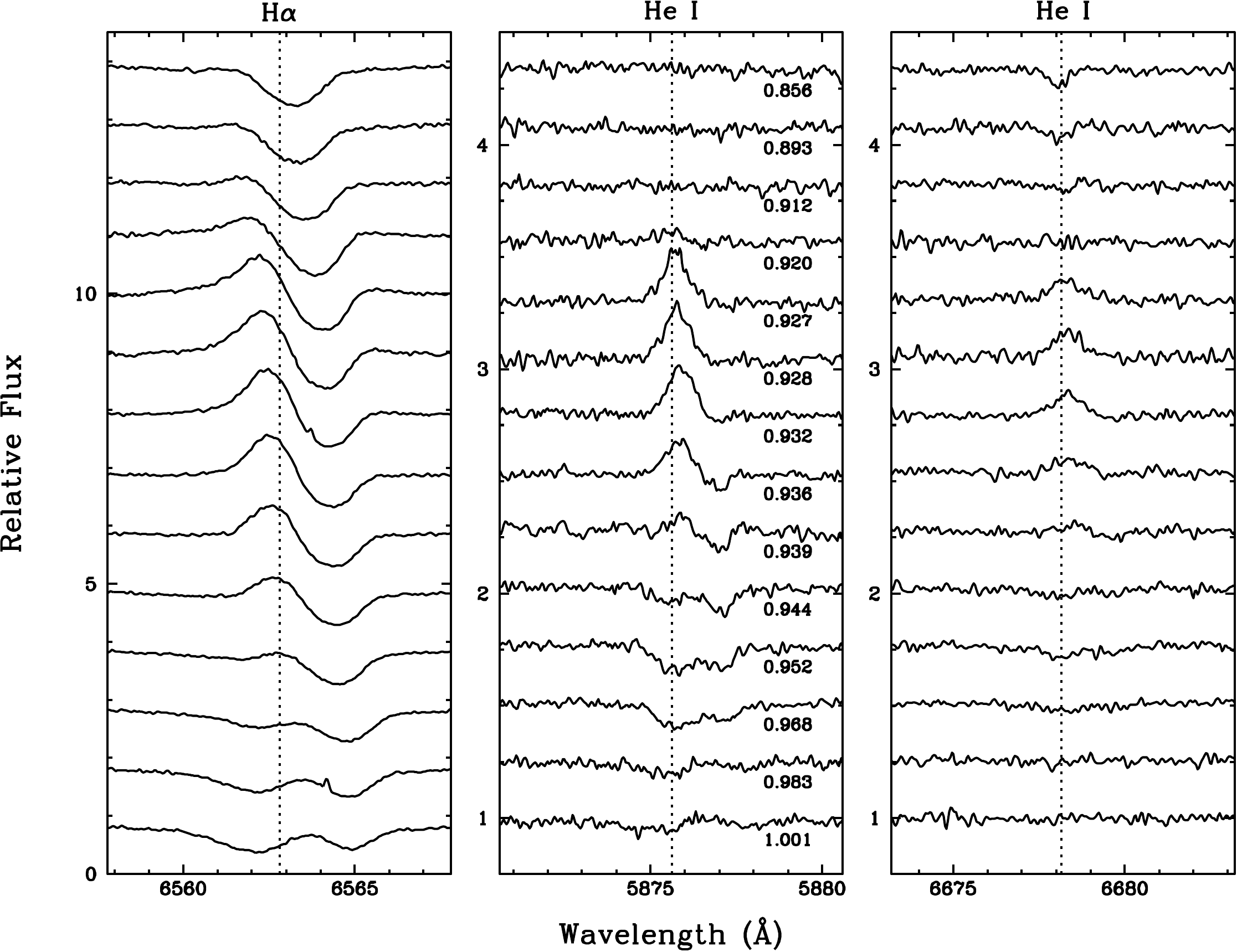}
\end{center}
\caption{\label{fig01}
\footnotesize
   Variations of H$\alpha$, \species{He}{i} 5875.62~\AA, and \species{He}{i} 
   6678.15~\AA, in the rising light shock phases of RV~Oct.  
   Dotted lines indicate the rest wavelengths of these lines as defined by
   the metallic lines.
   The phases are written in the middle panel.
}                                                               
\end{figure}

\begin{figure}
\begin{center}
\includegraphics[scale=0.65,angle=-90]{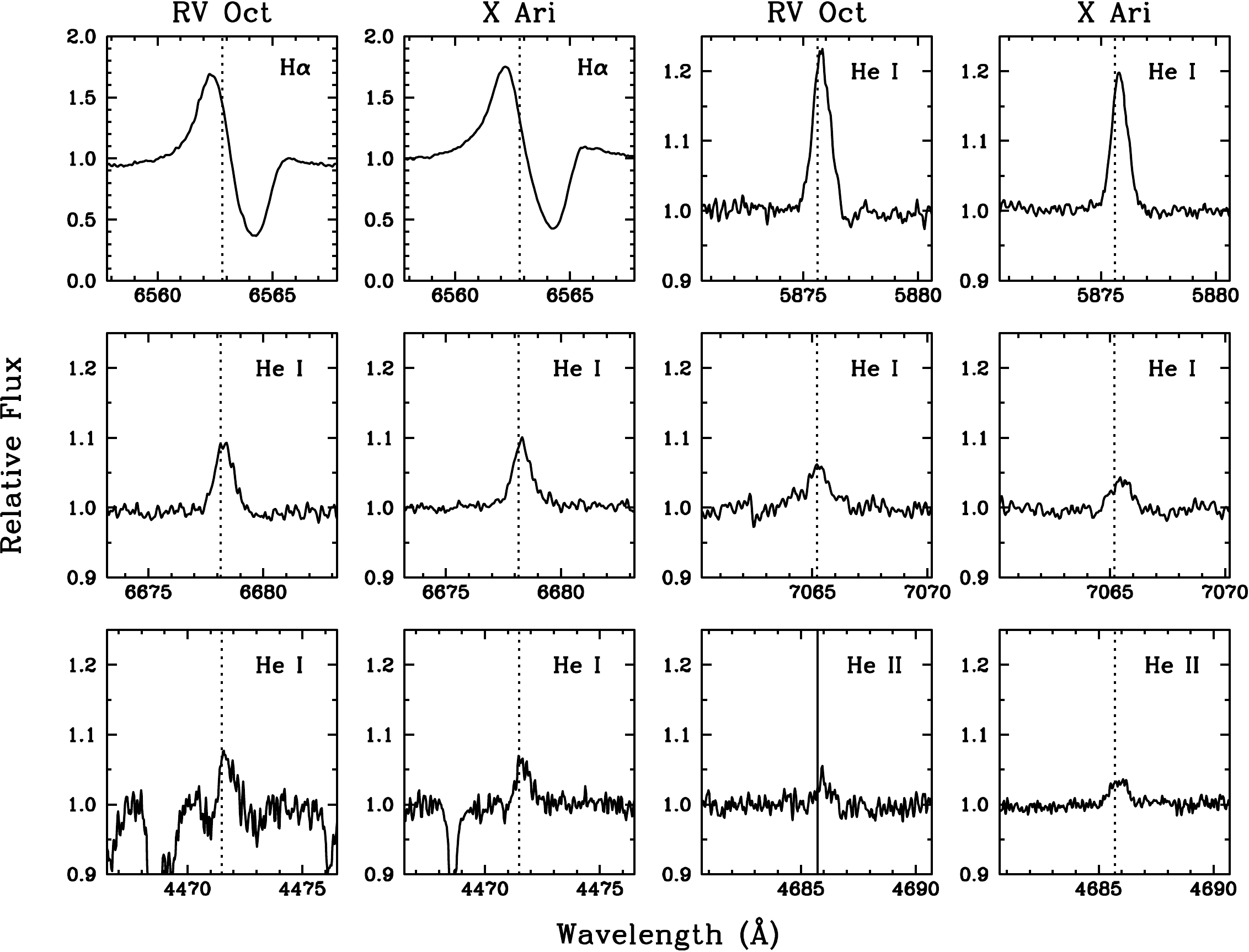}
\end{center}
\caption{\label{fig02}
\footnotesize
   Mean line profiles of H$\alpha$, four \species{He}{i} lines and a 
   \species{He}{ii} line at the phase of maximum He emission in program stars 
   RV~Oct and X~Ari.
}                                                               
\end{figure}

\subsection{Observations and Reductions}\label{obsreduc}

For this study we mine the database of several thousand 
echelle spectra of RR~Lyrae stars that were obtained with the du Pont 2.5~m 
telescope at Las Campanas Observatory between 2006 and 2014.
The resolving power of these spectra is $R \equiv\ \lambda/\delta\lambda$
=~27,000 at 5000~\AA.
We concentrate on a subset of stars analyzed by \cite{chadid17} that have
many spectra in the phase range $\phi$~=~0.8$-$1.1.
In Table~\ref{tab-stars} we list the observed photometric 
pulsational properties and
metallicities of the RR~Lyraes considered in the present work.
The pulsational data are taken from \citeauthor{chadid17} Table~1.
The adopted metallicities are means of the [\species{Fe}{i}/H] and 
[\species{Fe}{ii}/H] values (with assumed log~$\epsilon$(Fe)$_\odot$~=~7.50) 
listed in Table~2 of that paper.

RR  Lyrae stars pulsate with periods mostly confined to the range 
0.25~d $<$ P $<$ 0.75~d.  
Target exposure times, limited to small fractions of these periods, never 
exceeded 600~s ($\sim$0.01P), which resulted in relatively 
poor signal-to-noise ratios in the blue spectral region, S/N~$\sim$~15$-$20 
for most of the stars in our sample.
This created challenges for derivation of photospheric quantities (\teff,
\logg, \vmicro, [Fe/H] metallicity, [X/Fe] abundance ratios) from these 
spectra, because the great majority of  detectable metallic lines in RR~Lyrae 
stars occur at wavelengths less than 5000~\AA.
On the other hand, the H and He transitions used in this investigation lie 
in the yellow and red region, $\lambda$~$>$~5000~\AA, where our spectra have 
much higher $S/N$~$\sim$~50$-$80.

The du Pont echelle observations were reduced to final spectra as described 
in detail by \cite{for11a}.
These spectra have been used in multiple papers since that time. 
Therefore the reduction procedures will not be repeated here.
The individual du Pont multi-order wavelength calibrated spectra of all stars 
that we obtained in the 2006-2014 time period, whether or not they have
been used in the present paper, can be obtained at the Zenodo data
depository\footnote{
https://zenodo.org/record/5794389\#.YcD3XBPMIo8}

\subsection{The Transitions}\label{transitions}

In Figure~\ref{fig01} we extend Figure~1 of \cite{preston09} by showing 
the variation with phase near maximum light of RV~Oct for H$\alpha$ and the 
two \species{He}{i} lines with strongest emission, 5875.62~\AA\ and 6678.15~\AA.
The rapid appearance and fadeout of \species{He}{i} emission lines are
illustrated in the sequence of RV~Oct spectra in the phase range
0.66~$\lesssim$~$\phi$~$\lesssim$~1.00.  
Discussion of detailed H$\alpha$ phase variations has been included in
previous papers in this series for RRab stars \citep{chadid17} and
for RRc \citep{sneden17}.
For RV~Oct the \species{He}{i} emission occurs only in the narrow phase
range $\phi$~=~0.92$-$0.94, and we will show in \S\ref{helium1} that a 
similarly narrow phase range for He emission occur in all of our RRab stars.

Inspection of the \species{He}{i} lines for the three earliest phases
($\phi$~= 0.856, 0.893, 0.912) in the right-had panels of 
Figure~\ref{fig01} reveals apparent He absorption in the 6678~\AA\ line, 
but neither emission nor absorption in the 5876~\AA\ line.
The absorption seen at 6678~\AA\ throughout most of the pulsation cycle is,
in fact, due to \species{Fe}{i} not \species{He}{i}.
Most lines of \species{Fe}{i} in the red spectral region of \species{Fe}{i}
are very weak, but multiplet 268 \citep{moore72} is the exception with three
strong lines: 6546.25, 6592.92, and 6677.99~\AA.
These lines have similar transition probabilities \citep{kramida19}\footnote{
https://physics.nist.gov/PhysRefData/ASD/lines\_form.html}
and have substantial absorptions in the solar spectrum \citep{moore66}.
For $\lambda$6546 log~$gf$~=~$-$1.54 and EW$_\odot$~= 103~m\AA, 
somewhat blended with a weak \species{Ti}{i} line; 
for $\lambda$6592 log~$gf$~=~$-$1.47 and EW$_\odot$~=~123~m\AA; and 
for $\lambda$6678 log~$gf$~= $-$1.42 and EW$_\odot$~=~122~m\AA.
In the RV~Oct phases shown in Figure~\ref{fig01} we always detected
the $\lambda$6592 line and tentatively identified $\lambda$6546 above the
continuum noise.
However, the maximum-light phases under study here also have the highest
photospheric temperatures.
For RV~Oct and other RRab stars, 
$\langle \teff\rangle$~$\sim$~6150~K at $\phi$~$\sim$~0.85 
but rises sharply in about an hour to 
$\langle \teff \rangle$~$\sim$~7150~K at $\phi$~$\sim$~0.95 \citep{for11b}.
Neutral-species transitions such as those of \species{Fe}{i} weaken
with rising temperature.
If we examine the RV~Oct spectrum at earlier phases, all three lines
are easily detected.
For our spectrum at $\phi$~=~0.370 (\teff~$\sim$~6100~K, \citealt{for11b}), 
the equivalent widths are 
$EW$~=~27~m\AA\ for $\lambda$6546,
$EW$~=~35~m\AA\ for $\lambda$6592, and
$EW$~=~44~m\AA\ for $\lambda$6678.
All of these lines weaken to $EW$~$\lesssim$~5~m\AA\ at $\phi$~$\sim$~0.9,
but our conclusion holds: the $\lambda$6678 absorption is due to 
\species{Fe}{i}, not \species{He}{i}.

\cite{preston09} showed spectra of the \species{He}{i} $\lambda$5876 in 
emission during the rising-light phases of RV~Oct, but also argued
that four other He emission lines could be seen in these phases.
In Figure~\ref{fig02} we show the presence of multiple \species{He}{i}
lines in RV~Oct and the very metal-poor X~Ari, and one \species{He}{ii} in
both stars as well.
To assist in these identifications, we co-added individual observations
with maximum $\lambda$5876 emission, five of them for RV~Oct and eight
for X~Ari, thus substantially increasing their $S/N$ values.
The spectra that were combined were obtained essentially at a single phase
for each star.
Qualitatively the emission strengths are approximately the same in both stars.

\begin{figure}
\begin{center}
\includegraphics[scale=0.65]{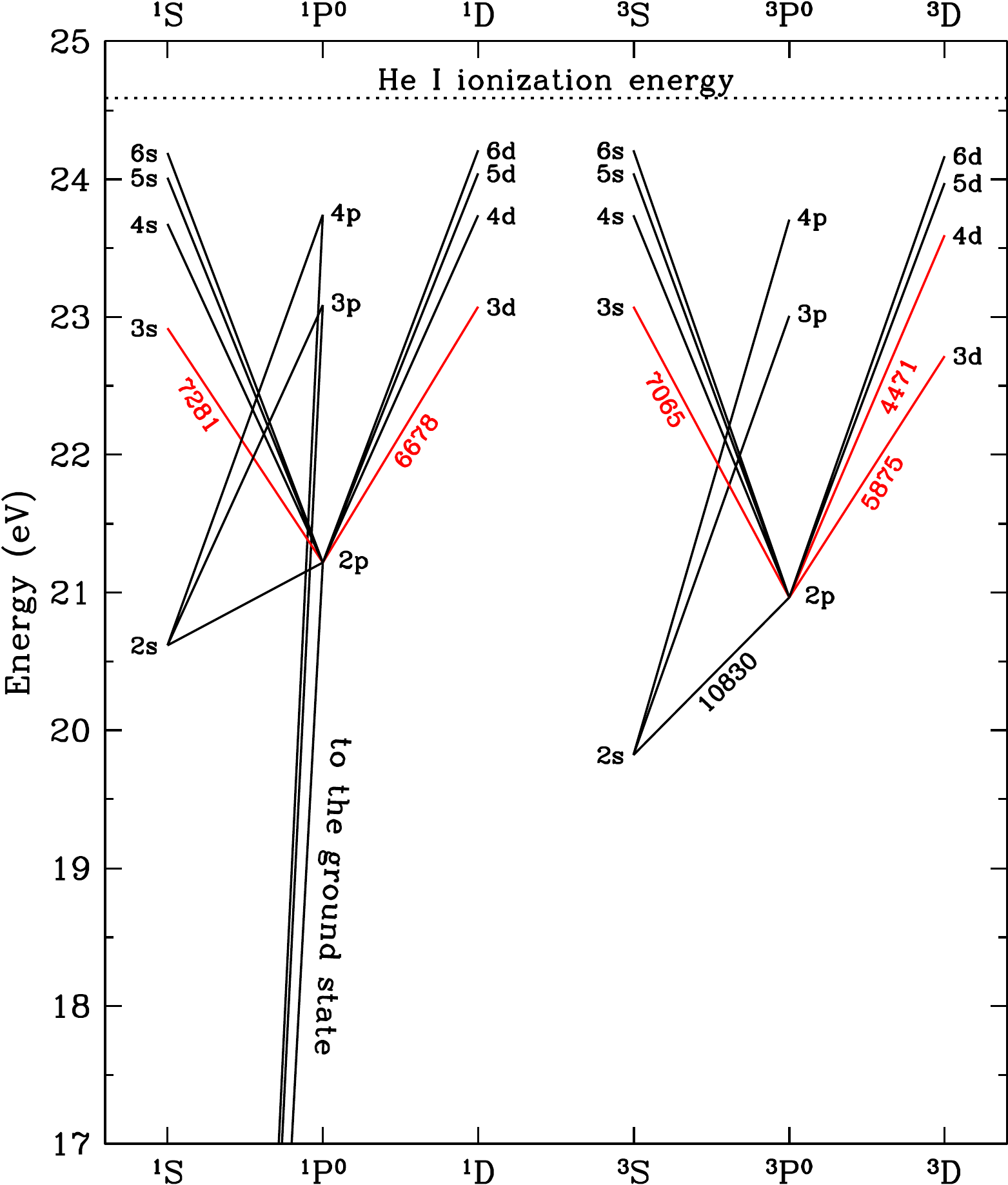}
\end{center}
\caption{\label{fig03}
\footnotesize
   A partial Grotrian diagram for \species{He}{i}.  
   Transitions detected by \cite{preston09} are in red, with the wavelengths
   labeled.
   No permitted dipole transitions connect the triplet and the ground states.
}
\end{figure}

The detected \species{He}{i} lines in RR~Lyraes arise from very high 
excitation states of two multiplet systems.
In Figure~\ref{fig03} we show a partial Grotrian diagram for this species,
modeled after \cite{moore68}, using data from the NIST atomic line
database \citep{kramida19}\footnote{
https://physics.nist.gov/PhysRefData/ASD/lines}.
The very high excitation energies of the 2$^1$P and 2$^3$P \species{He}{i} 
transitions ensure that these lines detected in RR~Lyrae stars cannot be 
formed in the same atmospheric layers that give rise to the metallic lines 
with typical excitation energies $\sim 1-3$~eV.

We omitted the 7281.48~\AA\ \species{He}{i} from Figure~\ref{fig02}.
This line is extremely weak in RV~Oct and not obviously present in X~Ari.
Additionally, the $\lambda$7280 spectral region is contaminated with many
telluric H$_2$O features that were not cancelled in our reduction procedures,
so we drop it from further consideration here. 
However, we should note that \cite{zhang05} found this transition most useful 
in constructing temperature diagnostics for planetary nebulae. 
We suggest that measures of \species{He}{i} $\lambda$7281 in spectra with 
S/N higher than ours may prove useful in analysis of RRab shocks as well.

\begin{figure}
\epsscale{0.60}
\plotone{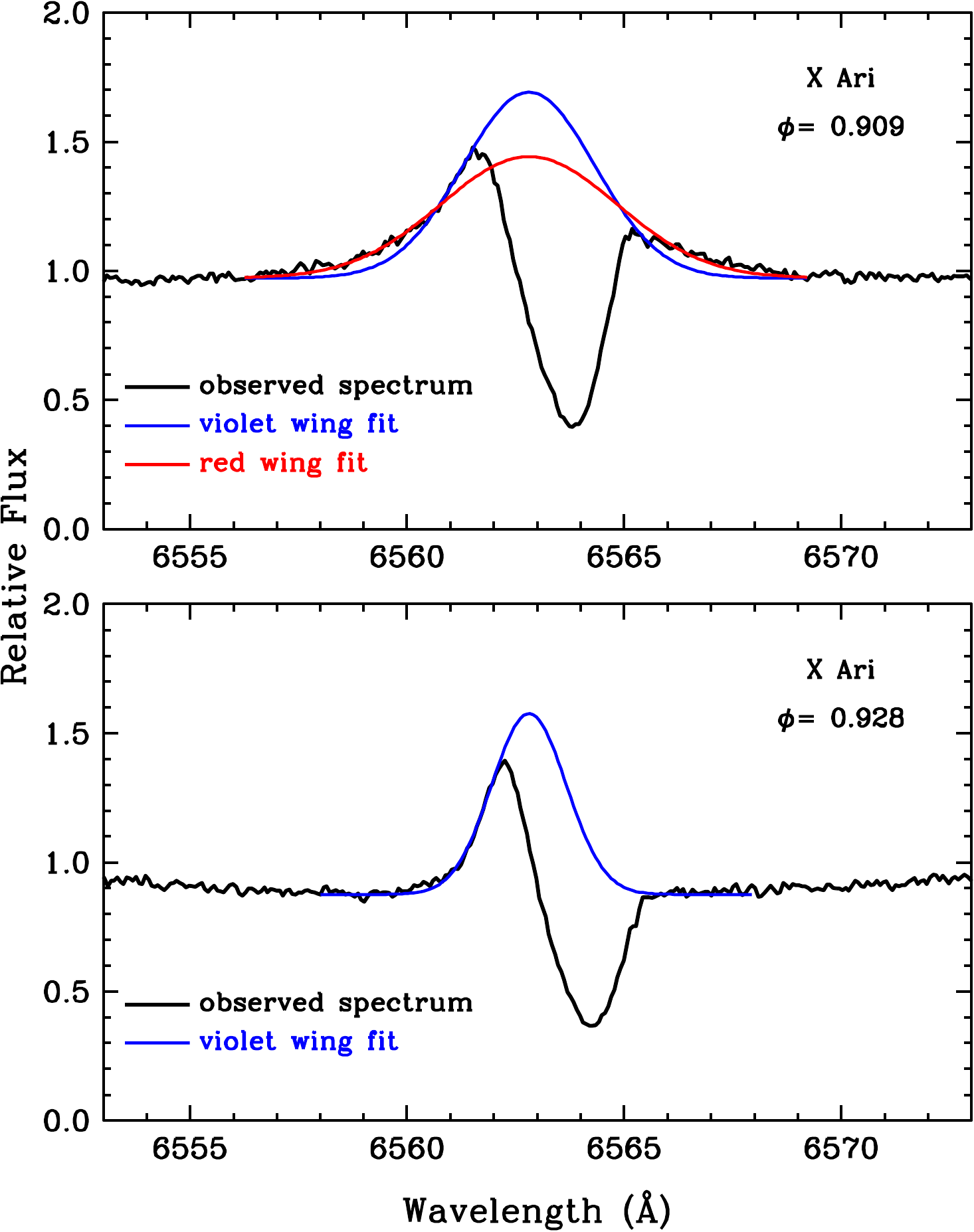}
\caption{\label{fig04}
\footnotesize
   Two representative H$\alpha$ lines near emission maximum.
   Top panel: the H$\alpha$ profile in X~Ari at phase $\phi$~=~0.909 in one 
   of its pulsational cycles (black line). 
   A Gaussian profile is fitted to the violet wing of the observed emission 
   (blue curve) ignoring the broad violet and red wings present in most stars 
   near emission maximum. 
   A second fit to the broad wings is plotted as a red curve.  
   Bottom panel: the same Gaussian fit (blue curve) is made to the observed
   H$\alpha$ profile for the more typical case at a later phase, 
   $\phi$~=~0.928, when there is no red emission component to constrain the
   Gaussian velocity center.
   Because of the general absence of this constraint, we adopted the 
   photospheric velocity defined by metal lines to locate the centroids of 
   all Gaussian fits.
}
\end{figure}

\vspace*{0.2in}
\section{MEASUREMENTS OF EMISSION AND ABSORPTION LINES}\label{measures}

\vspace*{0.2in}
\subsection{The Emission Lines}\label{emissions}

The onset of observable shock waves in RRab stars always begins with the 
sudden, simultaneous appearance of H and He emission lines.
These rise to maximum strength on time scales of minutes and decline on 
similar but slightly different timescales. 
They are followed by the appearance of violet- and red- displaced 
absorption components.

\subsubsection{H$\alpha$ Emission Equivalent Widths}\label{halphaflux}

We began with equivalent width (EW) measurements of the emission components of 
Balmer H$\alpha$ (6562.81~\AA).
Measurements of EWs and wavelengths of the Balmer lines are complicated 
by the fact that the emission profiles are overlain (thus partially masked) 
by strong red-shifted absorptions of gas infalling from the previous 
pulsation cycle.  

We measured the H$\alpha$ emission EWs in two ways: first 
by direct integration of emission parts of the observed features, and second
by modeling the Gaussian pure emission profiles that best match the blue 
wings of the observed profiles.
Our procedures are illustrated in Figure~\ref{fig04} for program star
X~Ari at two phases near maximum emission.

\begin{figure}
\epsscale{0.80}
\plotone{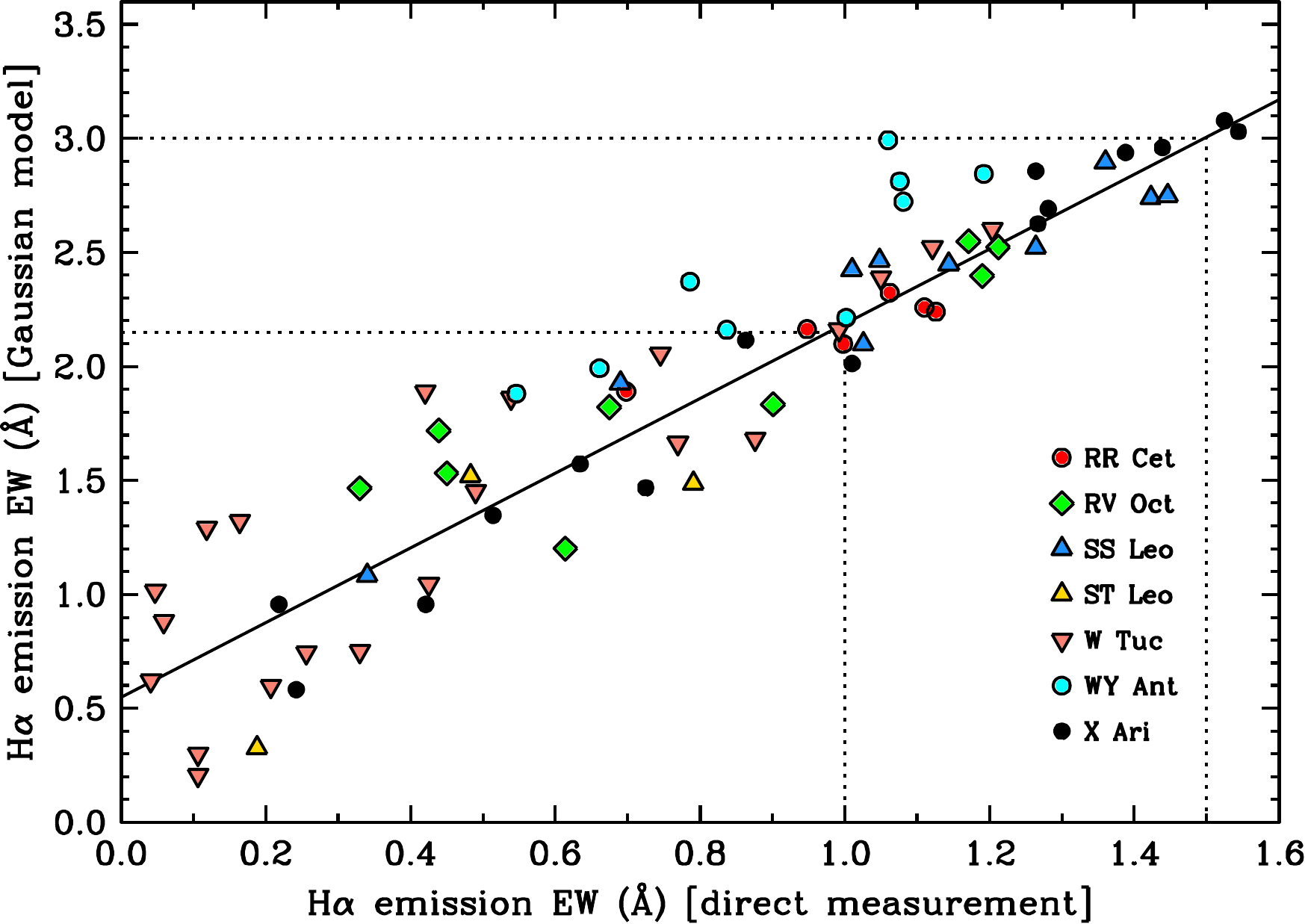}
\caption{\label{fig05}
\footnotesize
   The correlation of H$\alpha$ emission equivalent widths measured by
   direct integration and a Gaussian model approximation, as defined in
   the text.
   Data between dotted lines show that Gaussian model EWs exceed measured 
   values by a factor of 2.
}
\end{figure}

To effect direct integrations, we measured EWs of the violet 
emission fluxes in pixel ranges chosen by use of the IRAF 
\citep{tody86,tody93}\footnote{
http://ast.noao.edu/data/software}
$splot$ e-cursor option.
In Figure~\ref{fig04} for example, these ranges are 
$\sim$6557.0$-$6562.5~\AA\ (top panel) and 6560.0$-$6562.5~\AA\ 
(bottom panel).

For measurements of the Gaussian profiles we modeled the emission with 
Gaussians that best match the observed H$\alpha$ emission wings.
Following \cite{chadid13} we adopted the photospheric velocity defined by 
metal lines for the centers of our Gaussian fits.  
Then, as illustrated in Figure~\ref{fig04}, we were presented with
options for the best Gaussian approximations for the emission wings.
Inspection of the observed H$\alpha$ in the top panel reveals that a single 
Gaussian cannot be fitted to the entire emission part of the total profile.
In this case we fit a Gaussian to the violet component of X~Ari at 
phase $\phi$~=~0.909, ignoring the broad violet and red extensions, and a 
second fit to the broad extensions. 
The extended emission wings are a common, perhaps ubiquitous, characteristic 
of the H$\alpha$ emission profiles of RRab stars near emission maximum.  
They disappear at later phases, as illustrated in the Figure~\ref{fig04}
bottom panel for X~Ari at $\phi$~=~0.928 . 
The blue Gaussian emission wing had to suffice in the absence of a red emission
component.

\begin{figure}
\epsscale{0.80}
\plotone{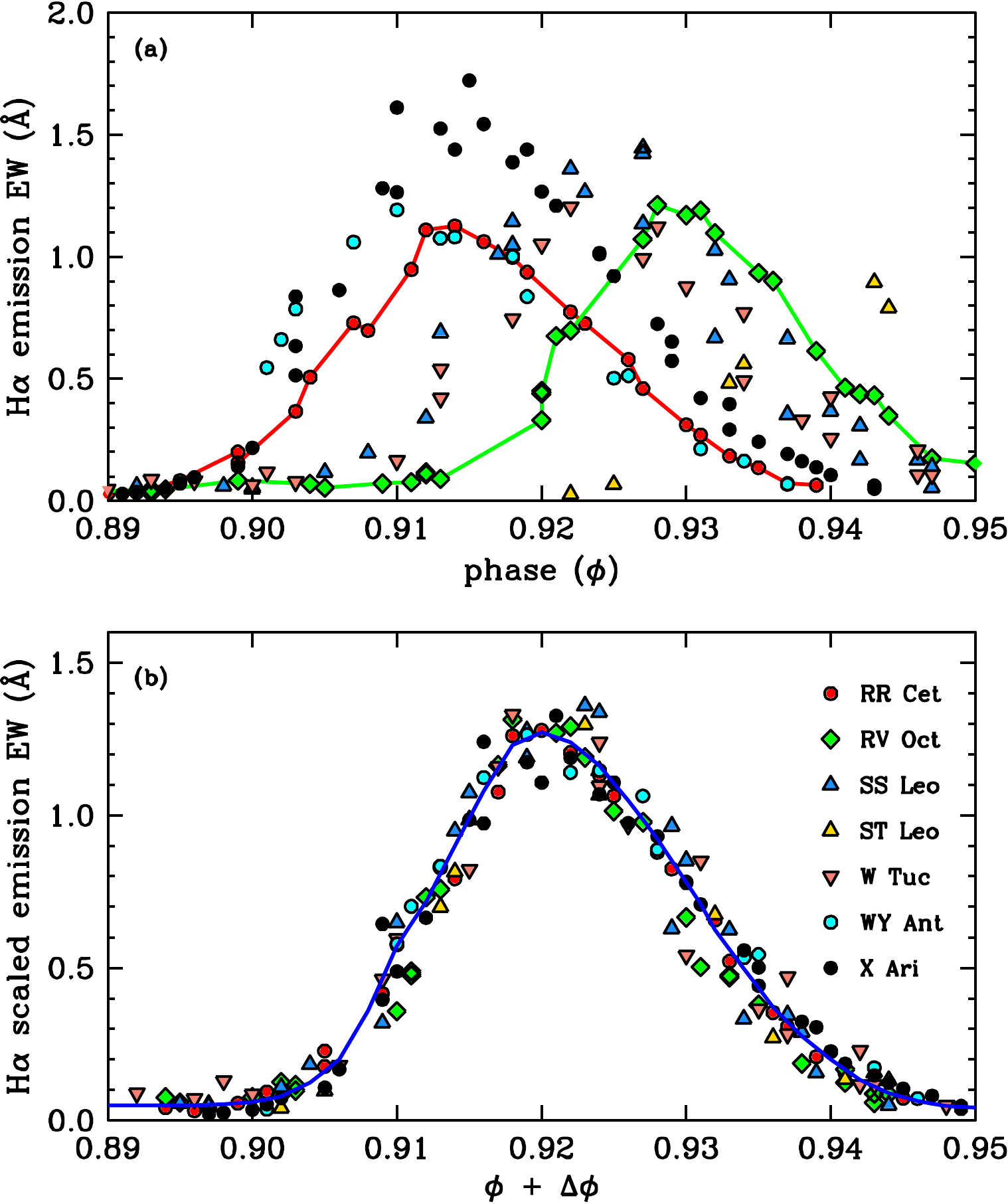}
\caption{\label{fig06}
\footnotesize
   Top panel: H$\alpha$ emission EWs measured via direct integration
   plotted versus phases.
   For each star $\phi$~=~0.00 was set to be the phase of minimum radial 
   velocity.
   The red points for RR~Cet and the green points for RV~Oct have been 
   connected with solid lines of their colors.
   Bottom panel: the emission EWs scaled to a common flux maximum
   and shifted to a common emission peak phase; see text for 
   descriptions of these H$\alpha$ emission manipulations.
   Here the blue line represents the mean trend of scaled emission versus
   shifted phase.
   The symbols and colors of points are identified in this panel's legend
   and are consistent with those of Figure~\ref{fig05}.
}
\end{figure}

We compare our emission EWs from direct measurement with those
derived from Gaussian model profiles in Figure\ref{fig05}.
The two sets are clearly correlated, but there are systematic offsets between 
measures of different stars. 
Compare for example those of WY~Ant (cyan-filled circles) and X~Ari 
(black solid circles) in this figure.  
However, for the bulk of strong emissions (equivalent widths $EW$~$>$~1~\AA)
the Gaussian model profile values exceed direct integration ones 
by a factor of $\sim$2, as shown by the solid line drawn through the data. 
The correlation is strongest for the strongest H$\alpha$ emission features, 
\ie, those contained between the dotted-lines in Figure~\ref{fig05}
(data for WY~Ant excepted).
Weaker emissions are more difficult to measure primarily because of 
uncertainty in proper continuum location. 
The systematic offsets and larger scatters in the regression for smaller 
EWs is evident in the figure.

The linear correlation between emission EWs derived from 
direct integration and from Gaussian model fits yields confidence that H$\alpha$
relative emission strengths can be derived from the observed line profiles.
For the remainder of this paper we will employ the EWs from direct integration
only, as these measurements involve fewer assumptions than those from
Gaussian integration.

\subsubsection{Phase variation of H$\alpha$ emission}\label{halphaphase}

The emission appears abruptly, rises to maximum on a timescale of a few 
minutes, and declines on a similar timescale, as illustrated in 
Figure~\ref{fig06} for seven RRab stars with the strongest emission 
in our sample.  
At first sight the data in the top panel seem to form a hopeless jumble.
But inspection quickly reveals that the jumble is caused largely by systematic 
star-to-star phase shifts. 
This is illustrated by lines connecting the data for RR~Cet (red) and 
RV~Oct (green).
The points for these two stars can be superposed in phase by applying mean
shifts of $+$0.0060 and $-$0.0095 to the RR~Cet and RV~Oct, respectively.  

\begin{figure}                                                
\epsscale{0.90}                                               
\plotone{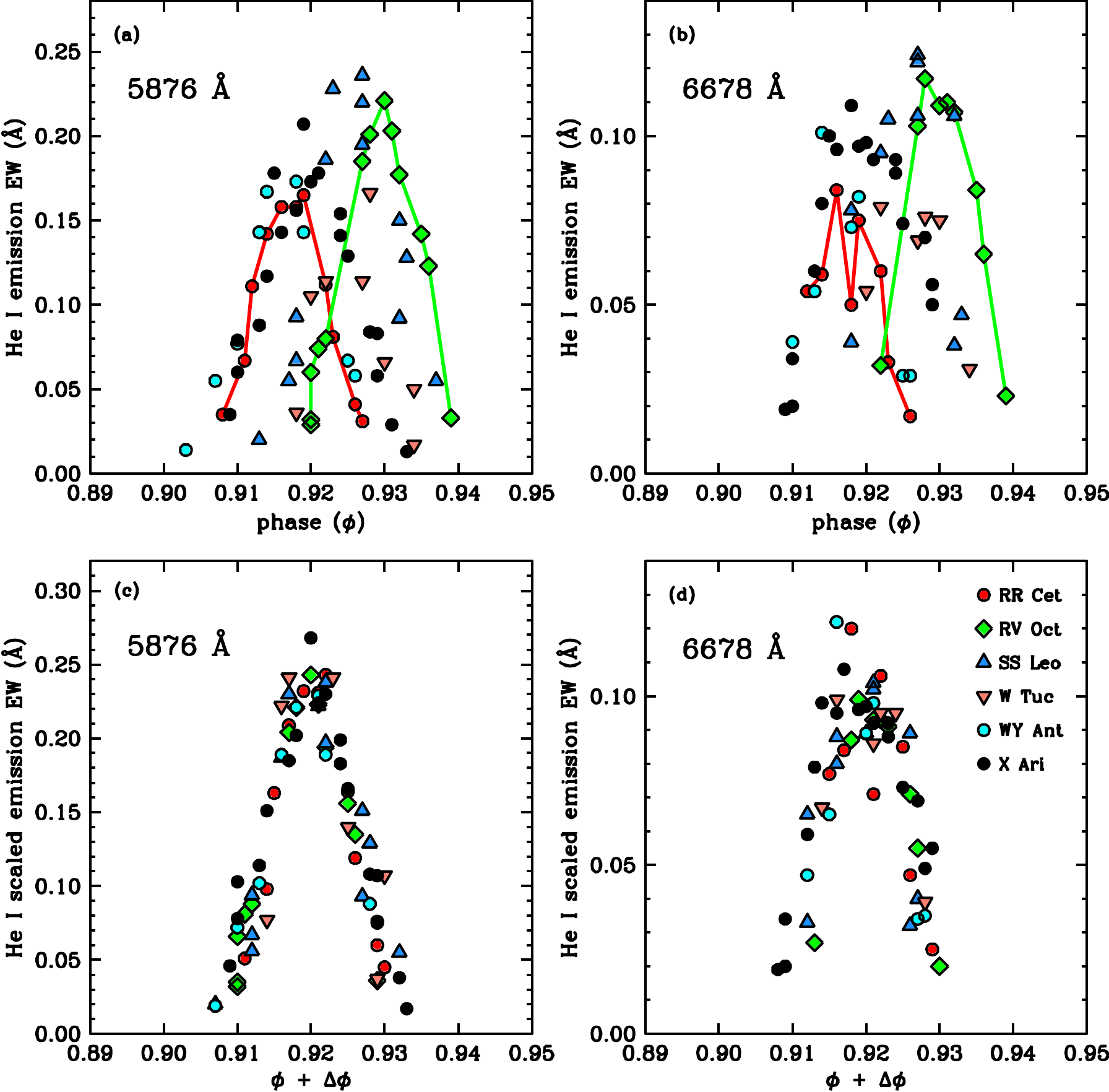}                                      
\caption{\label{fig07}                                   
\footnotesize                                                 
   Measured EWs (top panels) and scaled emission EWs shifted to 
   phase~=~0.92 (bottom panels) for \species{He}{i} 5876~\AA\ and 6678~\AA\ 
   in the manner employed for H$\alpha$ in Figure~\ref{fig06}.
   The symbols and colors of points are identified in panel (d)
   and are consistent with those of Figure~\ref{fig05}
}                                                             
\end{figure}

By trial and error we found phases shifts, $\Delta\phi$, that align all 
emission maxima at an arbitrarily adopted phase $\phi$~=~0.92.  
These phase shifts are listed in Table~\ref{tab-shifts}.  
Additionally, we used the absolute fluxes of \cite{kurucz11,kurucz18}\footnote{
Available at http://kurucz.harvard.edu/grids.html}
model atmospheres to convert measured fluxes in the du~Pont echelle orders 
to fluxes relative to the continuum flux at H$\alpha$, all scaled to the 
average value of these measurements in the phase interval 
0.915~$<$~$\phi\ + \Delta\phi$~$<$~0.925. 
We applied these flux normalizations and small phase shifts to the emission
phase data for the seven stars in the top panel of Figure~\ref{fig06} to 
produce the much more coherent flux versus phase variation in the bottom panel.
A blue-colored line in this panel represents the mean trend of these data.
The small scatter of individual points from this line, $\sim\pm$0.002 in
shifted phase, $\sim\pm$0.1 in scaled EW, strongly suggests that this is
a standard feature in H$\alpha$ behavior near maximum light in metal-poor
RR~Lyraes.

\subsubsection{Helium {\sc i} Emission}\label{helium1}

\begin{figure}
\epsscale{0.80}
\plotone{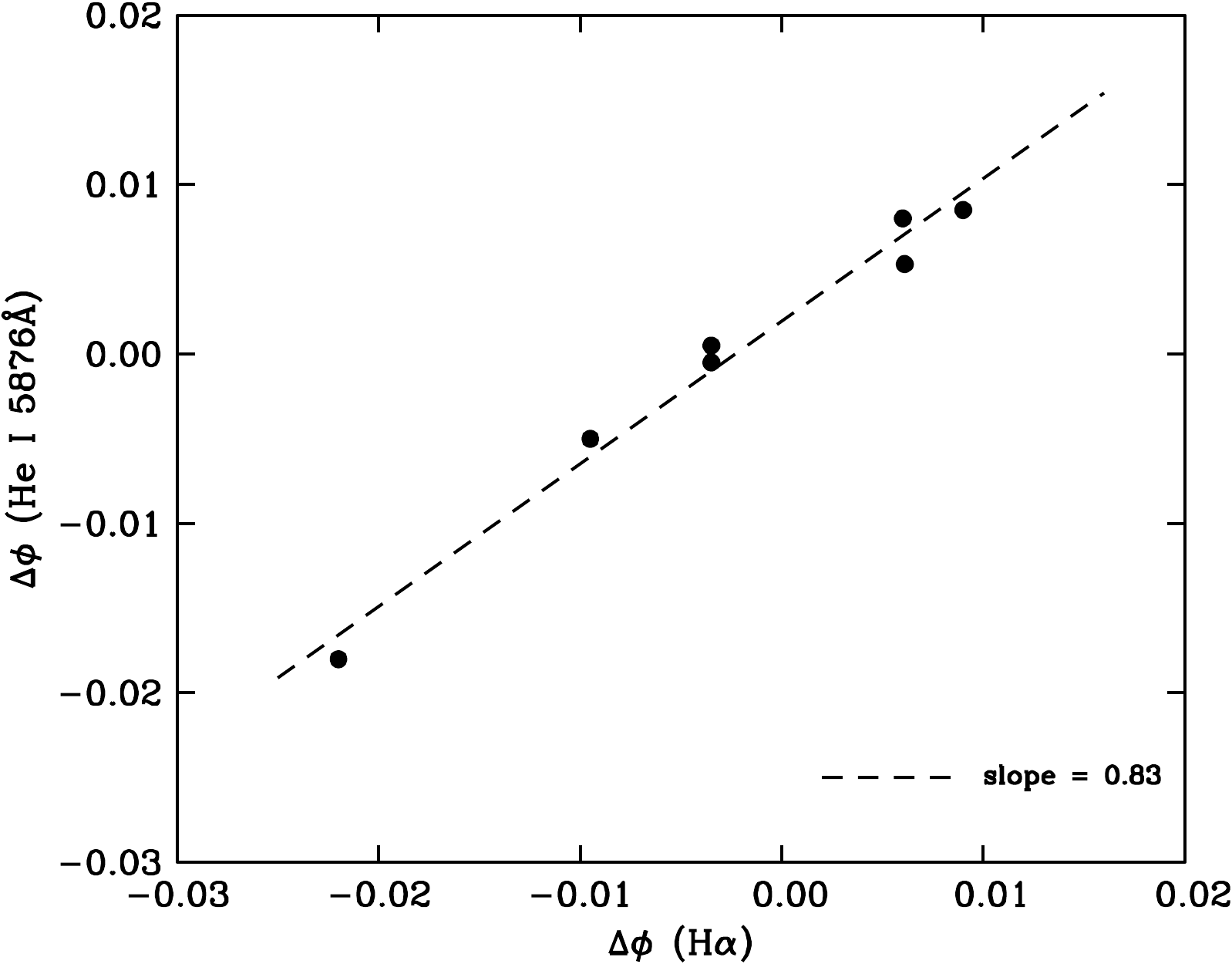}
\caption{\label{fig08}
\footnotesize
   Phases that produce alignment of emission maxima for \species{He}{i} 
   5876~\AA\ are plotted versus those for H$\alpha$. 
   The two shifts are strongly correlated
}
\end{figure}

Identifications of five emission lines of \species{He}{i} were reported by 
\cite{preston09} in du Pont echelle spectra of RV~Oct during rising light. 
We restrict attention to the leading lines of the series ($^3$P$^0$ $-$ $^3D$),
5876~\AA, and ($^1$P$^0$ $-$ $^1$D), 6678~\AA,  because they are the only
lines measurable in substantial numbers of the spectra obtained in our survey.
The behaviors of these two lines are much simpler than those of Balmer lines. 
Because of their much larger excitation energies ($>$20~eV), the 
\species{He}{i} series do not produce detectable absorption by infalling
gas during the initial emission line phases.
This simplicity is clearly illustrated in the middle panels of Figure~1 
(phases $\phi$ = 0.927, 0.928) above.
Therefore, flux and velocity measurements can be made without regard to the
complications encountered above for H$\alpha$.

Emission EW measurements of 5876~\AA\ and 6678~\AA\ were made in the same 
manner as described in \S\ref{halphaphase} for H$\alpha$.  
Peak EWs for these \species{He}{i} lines are smaller than that of 
H$\alpha$ by factors of 6 and 12, respectively, so measurements were limited 
to a smaller range of phases, and measurement accuracies were reduced 
accordingly.

In Table~\ref{tab-shifts} we list the estimated phase shifts 
$\Delta\phi$ for \species{He}{i} 5876~\AA\ needed to construct 
Figure~\ref{fig07} (b).
We emphasize that the phase shifts $\Delta\phi$ for \species{He}{i} 5876~\AA\ 
emissions were determined independently from those derived for H$\alpha$
described in \S\ref{halphaphase}.
Although both lines arise in the recombination wake of the shock, the two 
line forming regions cannot be identical because of the large 
difference in the ionization potentials ($\sim$11~eV) of \species{H}{i} and
\species{He}{i}.
Figure~\ref{fig08} shows that the two shifts are strongly 
correlated, but the slope of the regression is less than unity.

\vspace*{0.2in}
\subsubsection{Helium II $\lambda$4686 Emission}\label{he2em}

\begin{figure}
\epsscale{0.80}
\plotone{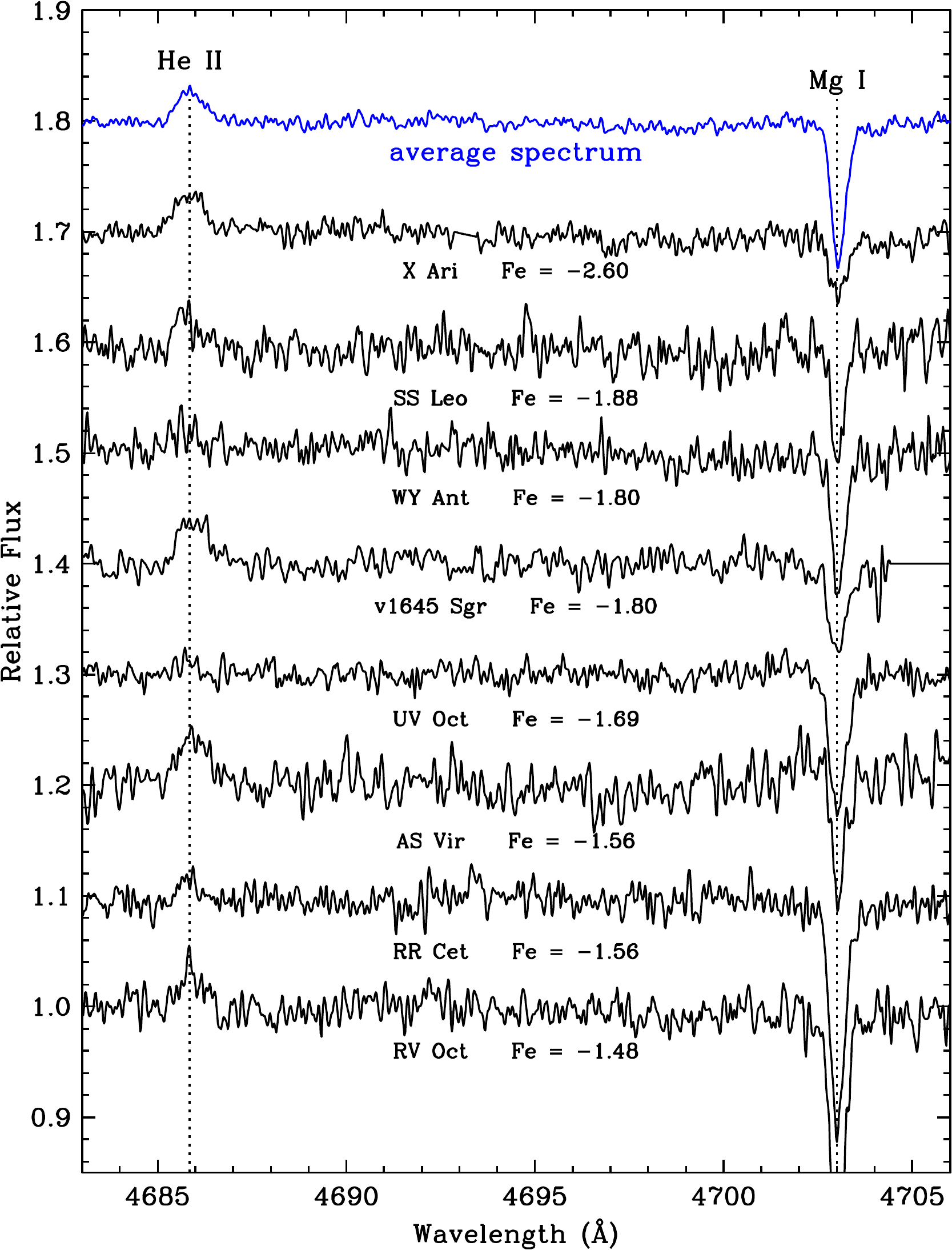}
\caption{\label{fig09}
\footnotesize
   Spectra of the \species{He}{ii} 4686~\AA\ region of eight 
   RRab stars arranged bottom to top in order of increasing [Fe/H].
   The relative flux values for RV~Oct are correct, and the spectra
   for the other stars are shifted vertically for display purposes.
   The average spectrum shown in blue at the top of the plot was formed by
   a straight mean of the indivdual spectra.
   Dotted vertical lines are located at the rest wavelengths of the 
   \species{He}{ii} line and the photospheric \species{Mg}{i} line.
}
\end{figure}
 
As noted in \S\ref{transitions}, \species{He}{ii} 4686~\AA\ 
3$^2$D$-$4$^2$F$^o$ is analogous to \species{H}{i} Paschen $\alpha$.
\cite{preston09} reported weak \species{He}{ii} 4686~\AA\ emission in three 
Blazhko variables (AS~Vir, UV~Oct, and V1645~Sgr) observed during large 
amplitude cycles when their metallic-line radial velocity amplitudes 
were 65, 70, and 64~\kmsec, respectively.\footnote{
Details of the radial velocity measurement procedure may be found in
\cite{preston00} and \cite{chadid17}.}
These emission lines together with those of five stable RRab with similar 
RV amplitudes are displayed in Figure~\ref{fig09}.
The average spectrum formed from these spectra at the top of the figure 
shows unambiguous detection of the \species{He}{ii} line.
The largest measured equivalent widths in Figure~\ref{fig09},
\ie, those for RV~Oct and~X Ari, are $\sim$30~m\AA.

By converting average Blazhko RV amplitude for AS~Vir, UV~Oct, and V1645~Sgr, 
66.3\kmsec, to pulsation amplitude, dR/dt\footnote{
We derive the pulsational velocity dR/dt in the stellar rest frame 
using formula (1) in \cite{chadid17}, as described in their \S4.3 .}
with projection factor, $p$ = 1.3, 
we obtain an estimate of a critical pulsation velocity amplitude, 
82~\kmsec, above which observable $\lambda$4686 \species{He}{ii}  emission 
occurs in our spectra.

We chose projection factor $p$ = 1.3 for calculation of dR/dt after review
of the $p$-values, derived by a variety of methods, that are summarized 
in Figure~5 of \citealt{neilson12}). 
These $p$-values, all sensitive to assumptions about limb darkening and 
atmospheric stratification, range from 1.2 to 1.4. 
We adopt a conservative one decimal-place value for $p$ that reflects our
uncertainties about RRab atmospheric parameters, particularly during 
rising light.

Finally, strong absorption due to \species{Mg}{i} at $\lambda$4703
illustrates the seemingly paradoxical occurrence of lines of 
\species{He}{ii} (IP~=~54.4~eV) and \species{Mg}{i} (IP~=~7.6~eV) in the 
same spectra. 
This circumstance requires pronounced temperature stratification in the 
shock wake, to be discussed in a work in progress.

\vspace*{0.2in}
\subsection{Absorption Lines}\label{absorptions}

Immediately after the phase of maximum emission, red-shifted absorption lines 
of H and He begin to appear in most metal-poor RRab spectra. 
This section is devoted to a description of their behavior.

\subsubsection{H$\alpha$ Absorption}\label{halphhaabs}

\begin{figure}
\epsscale{0.90}
\plotone{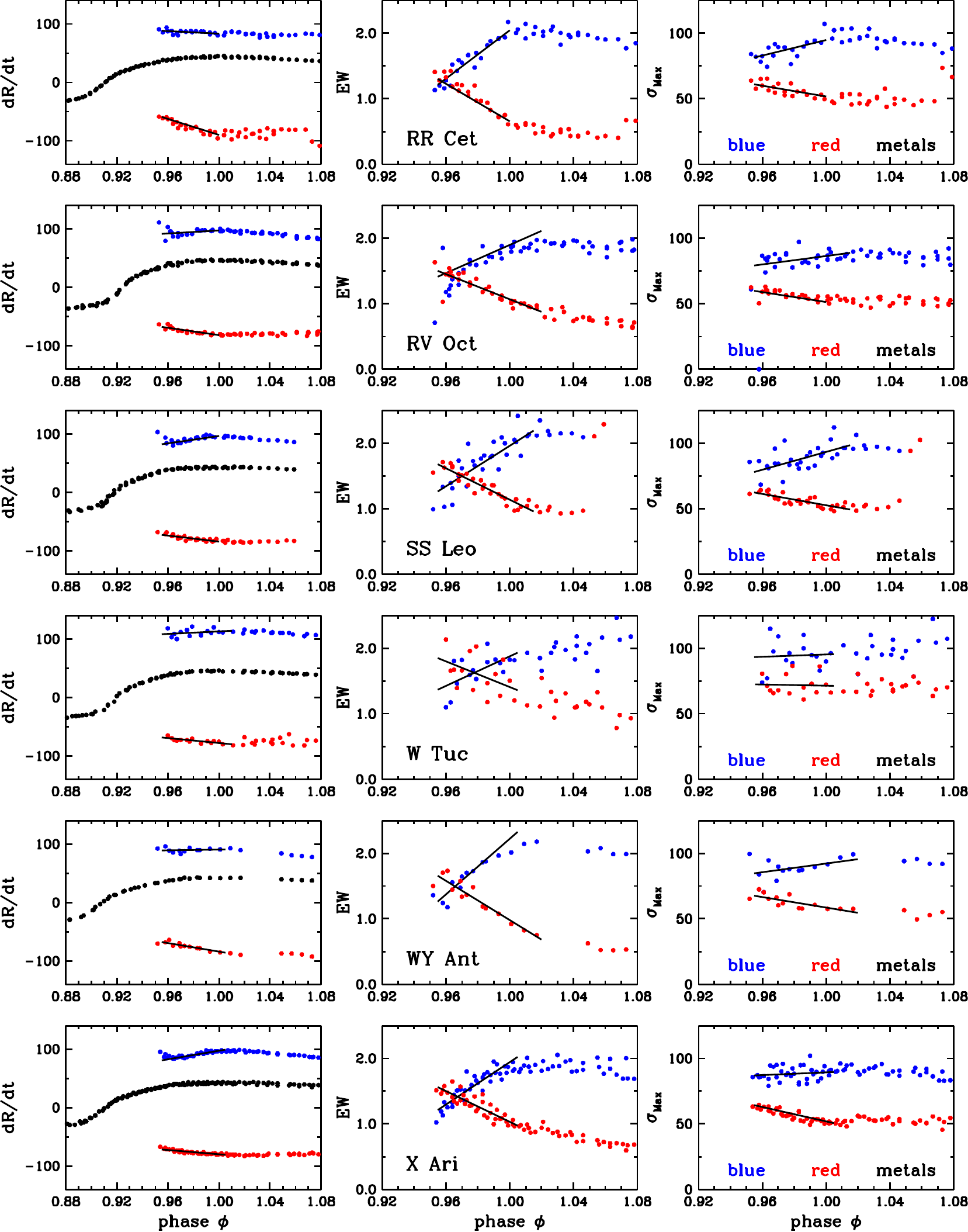}
\caption{\label{fig10}
\footnotesize
   The variations of H$\alpha$ pulsation velocity dR/dt (\kmsec), equivalent 
   width $EW$ (m\AA), and Maxwell line-of-sight velocity dispersion \sigmax\ 
   (\kmsec) with phase during rising light for six RRab stars.  
   Red and blue symbols are measures of lines produced by infalling and 
   out-flowing atmospheric layers, respectively. 
}
\end{figure}

We measured three quantities useful for discussion of 
atmospheric structure during rising light: radial velocities (RV \kmsec), 
equivalent widths (EW \AA), and Doppler widths (\sigmax\ \kmsec). 
To accomplish this task we used IRAF/splot with the ``d'' cursor option.
\sigmax\ is the Maxwellian line-of-sight velocity dispersion defined by 
equation~1 of \cite{unsold49}
We display the phase variations of these quantities for six RRab 
stars in the montage of Figure~\ref{fig10}.
Values of the pulsation velocities dR/dt and velocity dispersions 
\sigmax\ of the H$\alpha$ absorptions at phase $\sim$0.97, when they first 
become measurable, are given in the top half of Table~\ref{tab-kinrrab}.

The average H$\alpha$ velocity difference between infall and outflow in
column 4 of Table~\ref{tab-kinrrab} is about twice the range of the 
continuous velocity variation of the metal lines. 
Radial velocity is a location marker in the atmosphere.  
Metal lines are formed in the region where continuum optical depth is 
$\tau$~$\sim$~0.1$-$0.2, near the photosphere ($\tau$~$\sim$~$\frac{2}{3}$), 
while 
the heavily saturated cores of H$\alpha$ produced in the outflowing and 
infalling gas must be formed high in the atmosphere at very low continuum 
optical depths.  
The large persistent differences between H$\alpha$ and metal velocities when 
both are measured during rising light signal the existence of velocity 
gradients in the atmosphere detected by \cite{chadid96} and
discussed by \cite{bono94} and \cite{fokin97}.  
These gradients produce line asymmetries that will be the subject of a 
paper in preparation.

The regression coefficients used to construct the phase variations in 
Figure~\ref{fig10} for dR/dt and \sigmax\ are given in Table~\ref{tab-regress}.
We chose linear equations for all regressions with one exception: the large 
$S/N$ values of the \sigmax\ data for X~Ari permitted use of a quadratic fit.  
We see a tendency for maxima in \sigmax\ near phase 0.97 in the outflows of 
the other five stars, so we surmise that this quadratic behavior may well
be a general characteristic of the MP stars that is best studied with 
higher $S/N$ than that available in our spectra.

Two H$\alpha$ absorption components appear near phase 0.96 and remain 
measurable after maximum light for durations that vary greatly from star 
to star.  
The average difference between infall and outflow velocities, 
characteristically $\sim$165~\kmsec\ near maximum light, differs 
little from star to star in our sample. 
This similarity is due at least in part to a selection effect created by 
our choice of RRab stars with strong H$\alpha$ emission. 

The EWs of the outflow (B) and infall (R) components are approximately equal 
($\sim$1.5~\AA) at phase 0.97 for all six stars discussed here.  
The outflow values continue to increase for some 0.02 cycles past RV minimum.  
Thereafter, they remain constant or decline slightly.  
The infall EWs decline continually after their first 
appearance, more slowly after phase 0.02, and they generally persist 
beyond phase 0.10. 
Remarkably, infall absorption for RV~Oct can be measured to phase 0.2.

The line-of-sight velocity dispersions for H$\alpha$ are residual values 
after removal of instrumental broadening by Gaussian deconvolution, and much 
smaller effects due to thermal broadening, microturbulence and the Stark 
effect.
We discuss the (lack of) a significant contribution by the Stark effect to 
our measured line widths in \S\ref{stark}.

Outflow dispersions at phase 0.97 are typically $\sim$90~\kmsec, then 
they generally increase slightly until phase of RV minimum, and remain 
constant thereafter until the lines disappear. 
Infall dispersions at phase 0.97 are smaller ($\sim$62.0 $\pm$ 5.7~\kmsec), 
decline slowly and remain approximately constant at $\sim$50~\kmsec\
until they disappear. 
W~Tuc is an apparent exception.  
Our measures of its infall dispersion yield a constant value of 
72.0 $\pm$ 6.9~\kmsec\ in the phase interval 0.97~$<$~$\phi$~$<$~1.10.

\subsubsection{\species{He}{i} Absorption}\label{he1aabs}

\begin{figure}
\epsscale{0.90}
\plotone{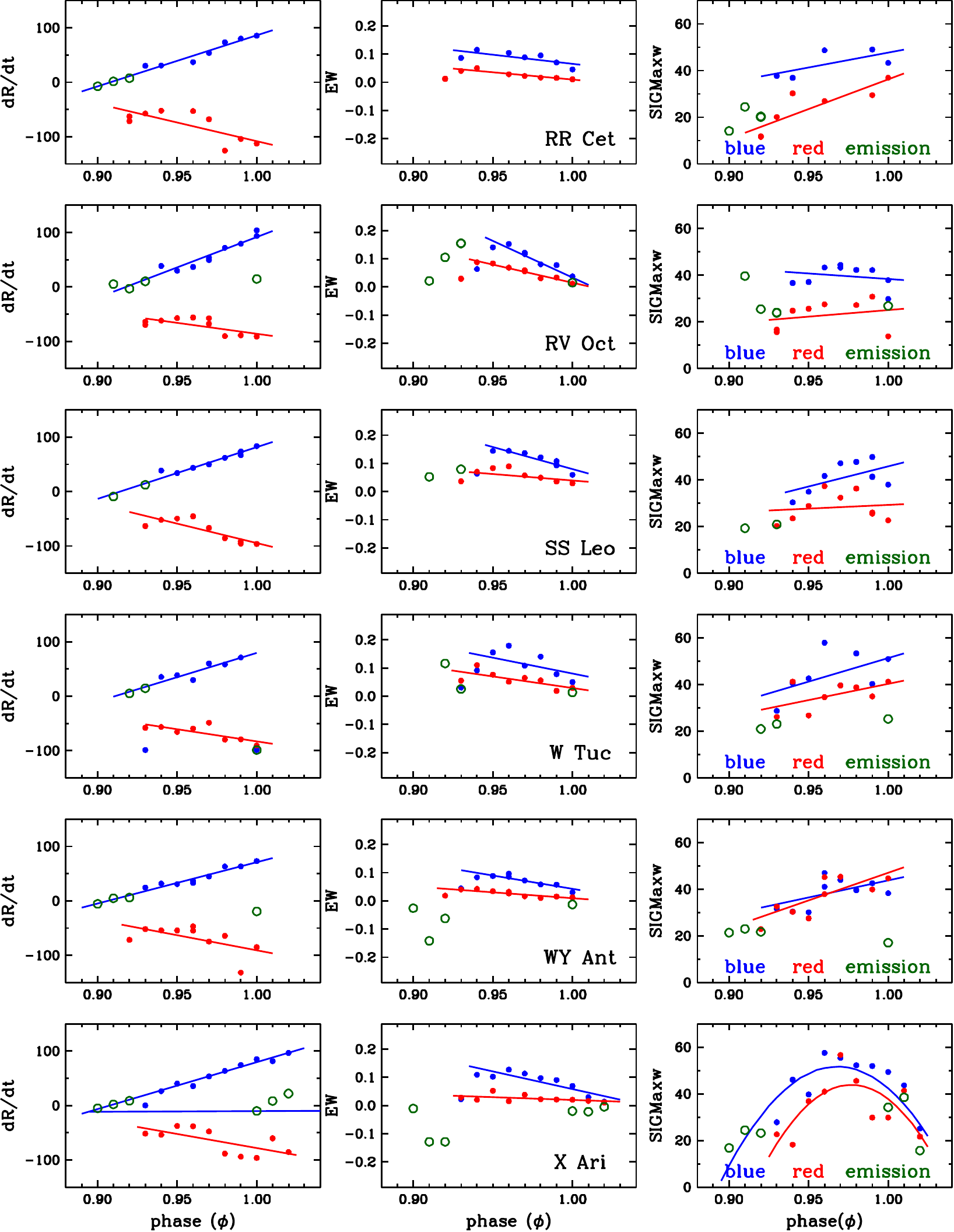}
\caption{\label{fig11}
\footnotesize
   (left) The variations of \species{He}{i} pulsation velocity dR/dt (\kmsec), 
   (middle) equivalent width $EW$ (m\AA), and (right) velocity dispersion 
   \sigmax\ (\kmsec) with phase during rising light for the same six RRab 
   stars shown in Figure~\ref{fig10}.  
   Red and blue symbols are measures of lines produced by infalling and 
   outflowing atmospheric layers, respectively. 
   Unfilled dark green symbols denote emission features.  
}
\end{figure}

\begin{figure}
\epsscale{0.60}
\includegraphics[scale=0.70,angle=-90]{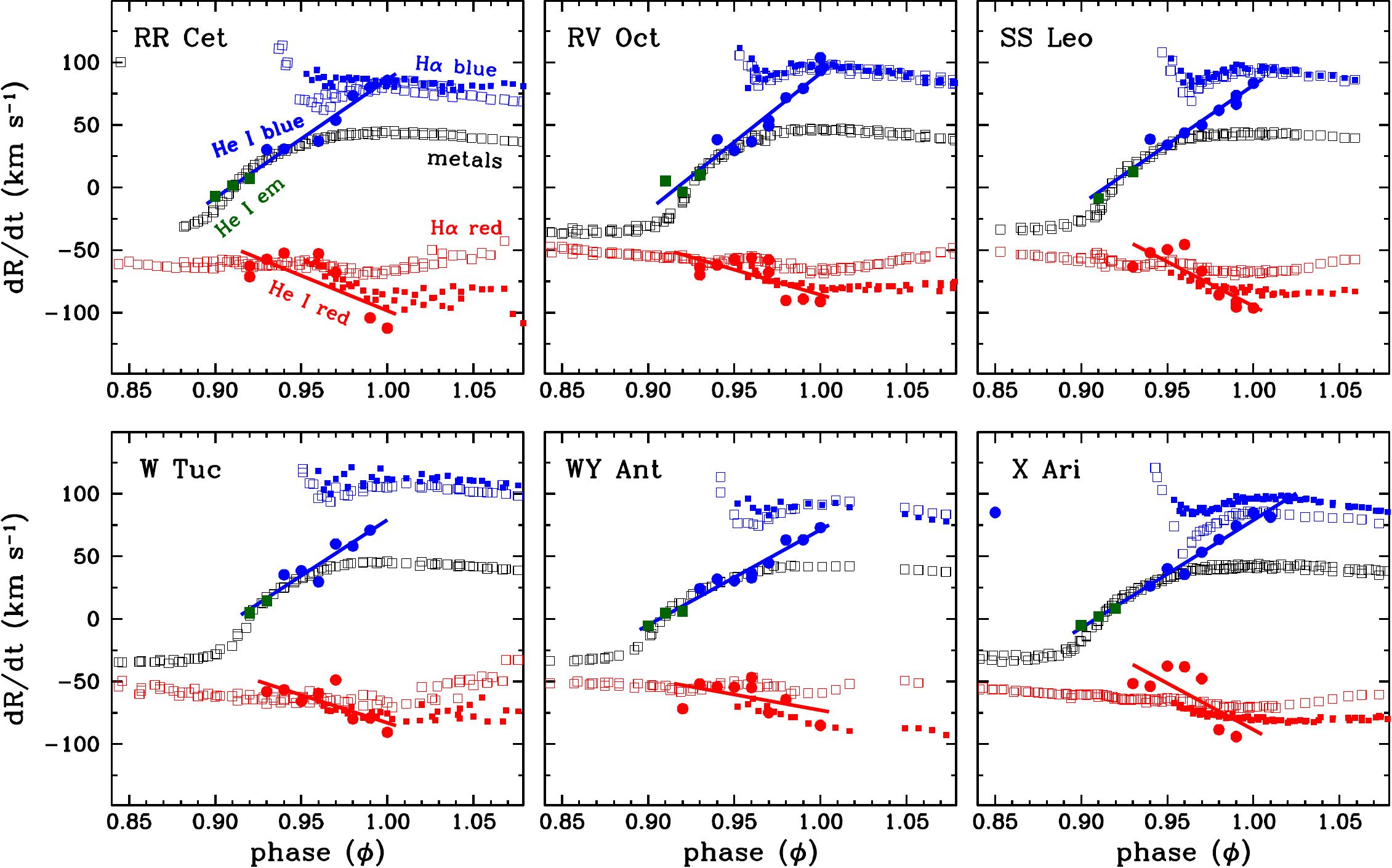}
\caption{\label{fig12}
\footnotesize
   H$\alpha$ and \species{He}{i} pulsation velocities from the left-hand 
   windows of Figures~\ref{fig10} and \ref{fig11} combined in 
   one panel each for the six stars.  
   The symbol colors are labeled in the panel for RR~Cet.
   Squares are used for the H$alpha$ data; open squares denote those 
   measurements made with IRAF $fxcor$, and filled squares for those made 
   with IRAF $splot$. 
   Data points for He 5876~\AA\ in this Figure represent more data than do 
   those in Figures~\ref{fig10} and \ref{fig11} because of co-addition. 
}
\end{figure}

\begin{figure}
\epsscale{0.90}
\plotone{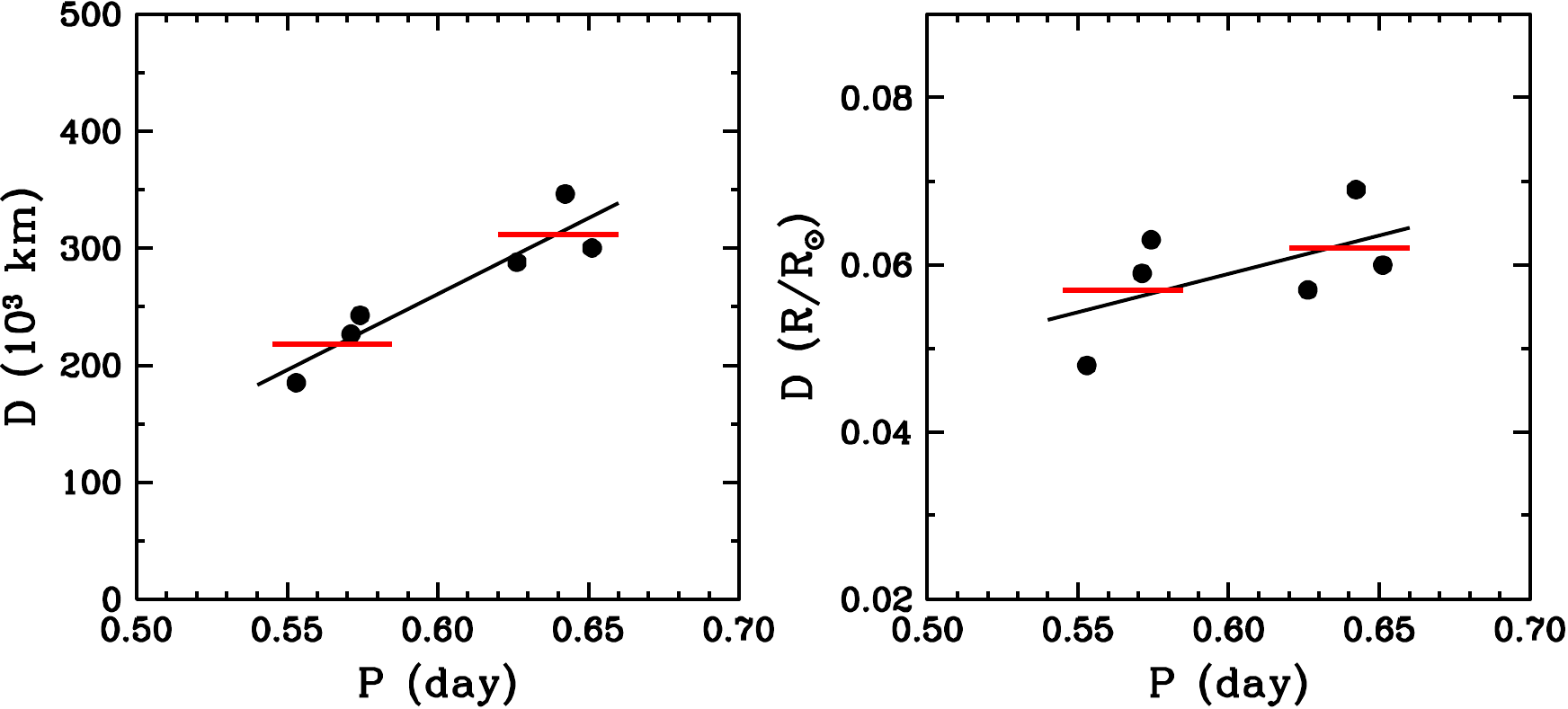}
\caption{\label{fig13}
\footnotesize
   The distance D traversed by the primary shock from the metallic line-forming 
   region just above photosphere to the layer that produces out-flowing 
   H$\alpha$ absorption.  
   In the left panel D is expressed in units of km, and in the right panel
   D is shown in units of stellar radii. 
   Data for six stars are taken from Table~\ref{fig13}. 
   Black lines show the mean trend of the data from a linear regression.
   Red lines show radii for Oo I and II groups taken from Table 1 of 
   \cite{marconi05}.
}
\end{figure}

We computed the same absorption quantities dR/dt (\kmsec), 
$EW$ (m\AA), and \sigmax\ that were derived for H$\alpha$.
These are displayed in Figure~\ref{fig11}.
Because \species{He}{i} 5876~\AA\ is much weaker than H$\alpha$ we co-added 
files into phase bins of 0.01P for this figure.
Emission features are included in the regression for dR/dt to illustrate 
the continuity with outflowing absorption that is expected, because the 
emission and blue absorption features must arise in the wake of the shock.

The small values of dR/dt for He $\lambda$5876, similar to those for the 
metal lines in the early expansion phases (0.90~$<$~$\phi$~$<$~0.95),
surprise us because the thermal conditions for the production of metal 
lines and He $\lambda$5876 are so different.  
Likewise, the velocity dispersion at these phases are also similar to those 
of the metals at these phases reported in considerable detail by 
\cite{preston19}. 
Note also the reappearance of He $\lambda$5876 emission in the post maximum 
spectra of RV~Oct, WY~Ant, and X~Ari, first reported for RV~Oct in 
\cite{preston09}.
This reappearance is discussed in \S\ref{hepersistence} below.

We interpret the He $\lambda$5876 data in a preliminary way by superposing it 
on the H$\alpha$ data for six stars in Figure~\ref{fig12}.
Integration of the He $\lambda$5876 regression from the inception of 
emission until it reaches the outflow velocity of H$\alpha$ produces 
an estimate of the distance traversed by the shock from the phase of 
emergence from the photosphere to its arrival at the level of formation of 
the H$\alpha$ absorption.  
The results of such integration for our six stars are shown in 
Figure~\ref{fig13}.

Although not important in the context of our investigation, we call attention 
to the systematic differences between H$\alpha$ velocities measured 
with IRAF/$splot$-$d$ and IRAF/$fxcor$ evident in Figure~\ref{fig12}. 
These differences are largest in the weak infalling (R) post-maximum 
components.  
$splot$-$d$ employs a deblending procedure that attempts to recover the 
(Gaussian) central wavelengths, EWs, and FWHMs of the B and R components by a 
best fit algorithm.  
The $fxcor$ measurements rely on the measurer's choice of pixels to include in 
the $fxcor$ sampling of the blended B and R components.  
This choice is difficult to make for the R component when it appears as a 
weak feature in the red wing of a strong B component.  
Accordingly, we use the $splot$-$d$ results in our discussion, including the 
$fxcor$ data in Figure~\ref{fig12} only as a warning to those who might 
contemplate similar measurements.

\subsubsection{Stark Broadening}\label{stark}

After correcting measured FWHMs for instrumental, thermal and microturbulent 
broadening by use of Gaussian subtraction we converted FWHM to Maxwellian 
line-of-sight velocity dispersion as described in \cite{preston19} and 
presented in Table~\ref{tab-kinrrab} and in the right-most panels of 
Figures~\ref{fig10} and \ref{fig11}.  
Before discussing these results we consider whether or not Stark 
broadening, a common effect in stellar lines of H and He, contributes to 
the widths of the Doppler cores of
H$\alpha$ and \species{He}{i} $\lambda$5876 in RRab atmospheres,
here taken to have characteristic values of \teff~=~7,000~K and 
N$_e$~=~10$^{14}$ cm$^{-3}$.

For H$\alpha$ \cite{kowalski17} calculate FWHM~=~0.14~\AA\ 
(\sigmax~=~5~\kmsec) for the parameters (\teff~=~10,000~K and 
N$_e$~=~10$^{14}$ cm$^{-3}$) that they deem appropriate for solar and stellar 
flares.  
This value agrees well with a three order-of-magnitude downward 
extrapolation of the regression of \cite{griem83} for T~=~20,000~K used by 
\cite{kielkopf14} as a fit to laboratory data collected in the range 
10$^{17}$ $<$ N$_e$ $<$ 10$^{21}$.  
We conclude that Stark broadening is an insignificant contributor to our 
\sigmax\ values for H$\alpha$ in MP RRab stars.

For \species{He}{i} $\lambda$5876~\AA\ we refer to the regression used by 
\cite{gigosos14} in their Figure 24b to fit data obtained in a number of 
laboratory experiments.  
A small justifiable extrapolation of their regression to stellar N$_e$ 
values indicates that the Stark effect produces a negligible broadening of 
this line at T~$<$~10,000~K, N$_e$~=~10$^{14}$ cm$^{-3}$.
 
Accordingly, in the absence of any additional broadening processes, we 
conclude that the \sigmax\ values in Table~\ref{tab-kinrrab} and 
Figures~\ref{fig10} and \ref{fig11} are reliable measures of line-of-sight 
velocity dispersions that occur in MP RRab atmospheres during primary light 
rise.

\vspace*{0.2in}                                               
\section{THE NATURE, ORIGIN AND PERSISTENCE OF HE LINES IN RRab}\label{hepersistence}
                                                              
\begin{figure}
\epsscale{0.90}
\includegraphics[scale=0.75,angle=-90]{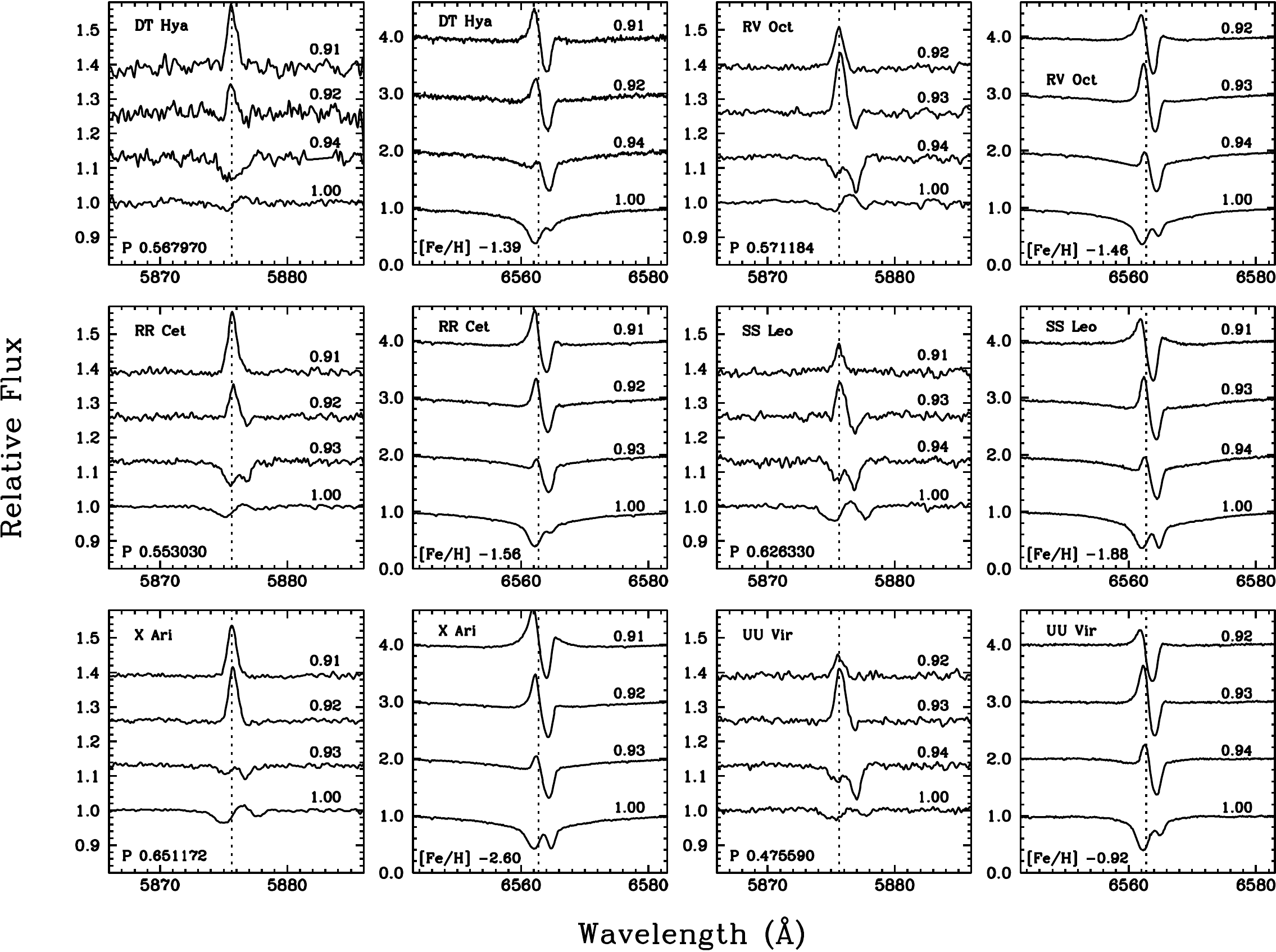}
\caption{\label{fig14}
\footnotesize
   Line profiles of \species{He}{i} 5876~\AA\ and H$\alpha$ for six 
   metal-poor RRab at four pulsation phases.
}
\end{figure}

\begin{figure}
\epsscale{0.90}
\includegraphics[scale=0.75,angle=-90]{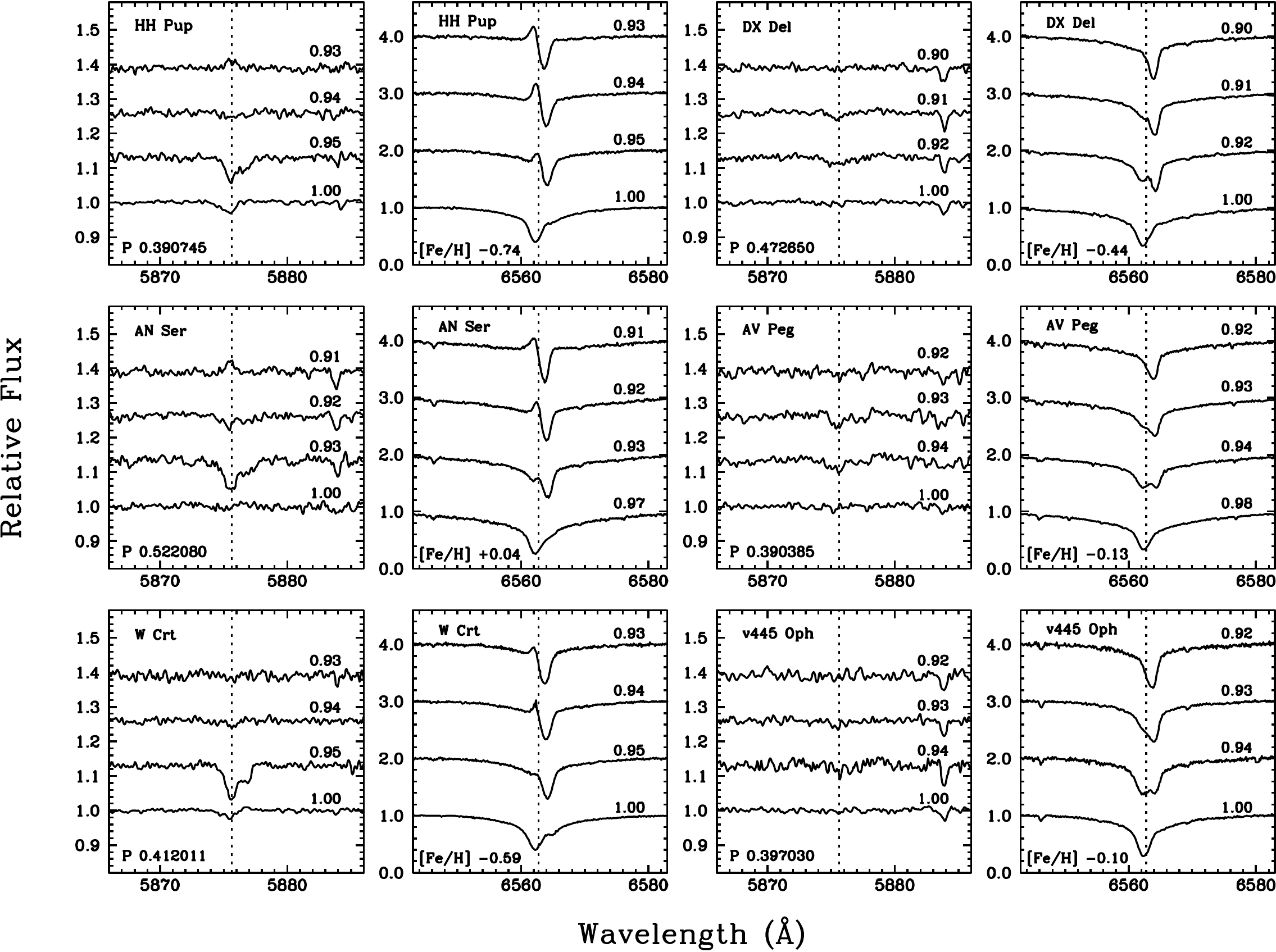}
\caption{\label{fig15}
\footnotesize
   Line profiles of \species{He}{i} 5876~\AA\ and H$\alpha$ for six 
   metal-rich RRab at four pulsation phases.
}
\end{figure}

Our discussion centers on Figures~\ref{fig14} and \ref{fig15}.
In these figures dashed vertical lines are plotted at the rest wavelengths 
of metal lines, which are formed in the near-photosphere gas layers.
We derived these rest wavelengths by cross-correlation of metallic-line 
spectra in the wavelength region 4000$-$4600~\AA, using IRAF/fxcor as 
discussed by \cite{preston00}.
Masks removed Balmer H$\delta$ and H$\gamma$, and the great majority of 
metallic absorption lines that create the cross-correlation signal are those 
of \species{Fe}{i} with excitation potentials between 1~eV and 3~eV.  
Dashed verticals at rest wavelengths closely bisect the \species{He}{i} 
5876~\AA\ emission features in all stars of both figures, but they lie 
slightly red-ward of the H$\alpha$ emission maxima as expected from the 
considerations in \S\ref{halphhaabs}.  
The weak \species{He}{i} lines in metal-rich stars conform to expectations 
based on H$\alpha$ in \cite{chadid17}.
However, time-evolution of the line profiles of \species{He}{i} 5876~\AA\ 
and H$\alpha$ in these figures pose several problems.

In previous investigations we divided our stars into metal-poor 
and metal-rich abundance groups at [Fe//H]~=~$-$1, rounding to the nearest 
integer the abundance boundary, $-$0.8, used by \cite{zinn85} to separate 
disk globular clusters from those of the halo.
Here we revert to Zinn's division point, primarily because it places 
UU~Vir among the metal-poor RRab, where it belongs in Figures~\ref{fig14} 
and \ref{fig15} and in the various correlations presented in 
\cite{chadid17}.
That such precision in choice of the division point affects our discussion 
strongly suggests, that the boundary is, in fact, blurred, \ie, the abundance 
distributions of halo and disk overlap. 
We reject the notion that UU Virginis is a class \textit{sui generis}.

Following expectations of \nocite{schwarzschild52}Schwarzchild's (1952) model, 
we locate the H and He emissions in outward-flowing gas following its passage 
through a strong shock \citep{wallerstein59b}. 
The different appearances of the initial emission line profiles of He and H 
arise as a circumstance imposed by infalling gas.  
H is so abundant that it produces absorption in atmospheres of all stars, 
types O through M.  
Infalling gas above the shocks will have temperatures 
that are consistent with those implied by broad-band colors and
metallic lines in the spectra (T~$\sim$~6500, \eg, \citealt{for11b}).
This gas is responsible for the strong red-ward displaced H$\alpha$ 
absorption.
Because \species{He}{i} lines, all with excitation potentials$>$20~eV 
(Figure~\ref{fig03}), occur only in OB atmospheres with \teff~$>$~10,000K, 
He emission flux produced in the wake of the shock is not distorted initially 
by absorption of infalling gas. 
Therefore $\lambda$5876 is an ideal diagnostic for investigating the 
time-evolution of shocks in RRab atmospheres.

However, issues arise with the evolution of He 5876 line profiles presented 
in the top three spectra for each star in Figure~\ref{fig14}. 
Stars in this Figure are arranged in order of decreasing maximum emission 
flux (from DT Hya upper left to UU Vir lower right).  
He absorption in the infalling gas appears in all six stars at or soon 
after phase 0.92. 
It strengthens initially as emission declines and it persists long after 
the observable emission disappears. 
This behavior is particularly noticeable in the panels for X Ari, RV Oct, 
SS Leo, and UU Vir. 

The origin of infalling \species{He}{i} 5876~\AA\ absorption presents a puzzle.
The p$^3$P$^o$ level of He $\lambda$5876 is metastable: radiative excitation 
from the 1S ground state to this state is forbidden, and in any event the 
lifetime of 2$^3$P$^o$ is a mere 5.63~ms (\citealt{lach01,hodgman09}). 
In contrast, prominent $\lambda$5876 absorption (EW~$\sim$~50~m\AA) is present 
in infalling gas throughout the phase interval 0.92~$<$~$\phi$~$<$~1.00, 
persisting for $\sim$1 hour after the phase of emission maximum as 
illustrated in Figures~\ref{fig11} and \ref{fig14}.

Collisional excitation cannot populate the 2$^3$P$^o$ level of infalling He 
because this gas has not yet reached the shock.  
So, how is this level populated?  
The only remaining possibility is recombination following ionization by a 
source of ``hot'' radiation. 
However, the broad-band colors of these stars during rising light of 
by \cite{monson17} provide no convincing evidence for such a long-lived 
source of hot radiation that would have to be present during all of the 
rising branch of the pulsation cycle and beyond.  

Next, we call attention to an unexpected change in the \species{He}{i} 
5876~\AA\ profiles that occurs during a few minutes near phase 0.97, 
intermediate between the phases of the bottom two spectra in each 
panel of Figure~\ref{fig14}.
Such variations of this line profile, first noted in spectra of RV~Oct 
\citep{preston09}, are reminiscent of the profile variations of 
\species{Fe}{ii} $\lambda$4923 encountered in the Blazhko variable S~Arae
by \cite{chadid08}, and attributed by them to a hypersonic shock that develops 
high in the atmosphere. 
Our \species{He}{i} phenomenon is seen most clearly by comparing the bottom 
two spectra of panels for X~Ari, RV~Oct, SS~Leo, and UU~Vir of
Figure~\ref{fig14}.
The symptoms of this phenomenon, present in varying degrees in all RRab 
stars in our survey, are these:
\begin{enumerate}
\item The profile suddenly broadens dramatically. 
The average radial velocity of the red-shifted absorption for the stars in
Figure~\ref{fig14} (DT~Hya excepted) increases from 60.3~$\pm$~3.7~\kmsec\
to 105.0~$\pm$~1.9~\kmsec\ while that for the violet-shifted component 
decreases from $-$10.5~$\pm$~8.0~\kmsec\ to $-$23.4~$\pm$~6.5~\kmsec. 
The extent to which the measured absorption velocities are shifted from 
their true positions by putative emission wings is unknown.
\item The velocity of the violet-shifted component is difficult to quantify 
unambiguously because of its asymmetry; the violet wing RV~Oct extends 
approximately to an escape velocity, $\sim$200~\kmsec, calculated for 
canonical values of the mass (0.65~M$_\odot$) and radius (5.5~R$_\odot$). 
\item Weak ($\sim$2\%) emission reappears between the absorption components. 
Chi-square analysis confirms its reality in X~Ari, RV~Oct, and SS~Leo.  
It is a reality to be reckoned with. 
\end{enumerate}

Finally, we consider the incongruent properties of two stars lying 
together at top-left in Figure~\ref{fig15}.
\begin{enumerate}
\item HH Pup, contrary to expectations due to its weak H and He emissions, 
has the largest photometric (1.35~mag) and radial velocity (69.7~\kmsec) 
amplitudes and the largest primary acceleration in our entire spectroscopic 
sample \citep{chadid17}.  
It also has the shortest pulsation period (0.39~d), which it shares with 
AV~Peg (also in the mid-right panels of Figure~\ref{fig15}). 
It is an archetypal example of the low latitude, short period RRab 
population first identified by \cite{kukarkin49} and subsequently as 
metal-rich by \cite{preston59}, \cite{layden94}, and \cite{layden96}.
\item AN~Ser, the most metal-rich star in our sample, [Fe/H]~=~+0.04, has by 
far the longest photometric period (0.52~d) of the metal-rich RRab. 
It belongs neither to Oosterhoff's type~I metal-poor cluster population, nor 
to Kukarkin’s short-period low-latitude metal-rich population.  
Note that DX~Del (P~=~0.47~d, [Fe/H]~=~$-$0.44) in the right column of 
Figure~\ref{fig15} is a cousin of HH Pup. 
\end{enumerate}

Location of HH~Pup, AN~Ser, and DX~Del in a proper evolutionary sequence of 
stellar populations is a challenge for the future. 
We continue to seek explanations of results presented above.
We defer our exploration of the systematically weaker H and He emission 
in metal-rich RRab to a paper in preparation.

\vspace*{0.2in}
\section{SUMMARY}\label{conclusions}

We report quantitative data about spectral features of H$\alpha$, 
\species{He}{i}, and \species{He}{ii} measurable with spectra obtained 
during  the course of an 8-year survey of some three dozen RRab stars with 
representative periods and metallic abundances. 
We present graphical and tabular descriptions of the time evolutions of the 
emission line equivalent widths of H$\alpha$, triplet \species{He}{i} 
$\lambda$5876, and singlet \species{He}{i} $\lambda$6678 of RRab stars with 
strongest \species{He}{i} emission. 
We also show time evolutions of the radial velocities, equivalent widths, 
and turbulent velocities of H$\alpha$ and \species{He}{i} $\lambda$5876 
absorptions measured for outflow and infall gas during the shock phases.

A principal conclusion of our work is that radial velocities of 
\species{He}{i} $\lambda$5876 define outward traveling waves in metal-poor 
RRab that we use to estimate the extent of RRab atmospheres traversed by the 
primary shock waves. 
We call attention to the problem of populating the 2$^3$P$^o$ level of 
\species{He}{i} in infalling gas during primary shock phases. 
We identify several RRab that do not fit in extant theory of 
Galactic stellar evolution. 
Finally we provide well-documented evidence for the occurrence of 
\species{He}{ii} $\lambda$4686 emission in eight RRab stars during primary 
light rise.

\begin{acknowledgments}

As always we thank the support staff for their efficient, 
cheerful assistance in our work at LCO. 
We also thank Giuseppe Bono and Harriet Dinerstein for helpful comments 
that improved our presentation.
This work has been supported by NSF grant AST-1616040 (C.S.).

\end{acknowledgments}

\facilities{Du Pont (echelle spectrograph)}

\software{IRAF (Tody 1986, Tody 1993),
SPECTRE (Fitzpatrick \& Sneden 1987)}


\clearpage
\begin{center}
\begin{deluxetable}{lrrrrr}
\tabletypesize{\footnotesize}
\tablewidth{0pt}
\tablecaption{Program Star Data\label{tab-stars}}
\tablecolumns{6}
\tablehead{
\colhead{Name}                     &
\colhead{$P$}                      &
\colhead{$T_0$}                    &
\colhead{$V$}                      &
\colhead{$V_{amp}$}                &
\colhead{[Fe/H]}                   \\
\colhead{}                         &
\colhead{(d)}                      &
\colhead{(d)}                      &
\colhead{(mag)}                    &
\colhead{(mag)}                    &
\colhead{}                         
}
\startdata
WY Ant    & 0.574340 & 1870.76 & 10.37 & 0.85 & $-$1.80 \\
X  Ari    & 0.651172 & 1890.10 &  9.24 & 0.94 & $-$2.60 \\
RR Cet    & 0.553030 & 2143.63 &  9.26 & 0.82 & $-$1.56 \\
W  Crt    & 0.412011 & 1871.64 & 10.90 & 1.10 & $-$0.59 \\
DX Del    & 0.472650 & 2415.85 &  9.81 & 0.70 & $-$0.44 \\
DT Hya    & 0.567970 & 1872.11 & 12.53 & 0.98 & $-$1.39 \\
SS Leo    & 0.626330 & 1873.06 & 10.47 & 1.00 & $-$1.88 \\
ST Leo    & 0.477983 & 2384.94 & 10.99 & 1.19 & $-$1.30 \\
RV Oct    & 0.571184 & 1891.66 & 10.53 & 1.13 & $-$1.46 \\
V445 Oph  & 0.397030 & 1939.03 & 10.54 & 0.81 & $-$0.10 \\
AV Peg    & 0.390385 & 2758.11 &  9.95 & 0.96 & $-$0.13 \\
HH Pup    & 0.390745 & 1869.65 & 10.57 & 1.24 & $-$0.74 \\
AN Ser    & 0.522080 & 2701.17 & 10.46 & 1.01 & $+$0.04 \\
W  Tuc    & 0.642260 & 1869.52 & 10.90 & 1.11 & $-$1.76 \\
UU Vir    & 0.475597 & 1886.54 & 10.06 & 1.08 & $-$0.92 \\
\enddata
\end{deluxetable}                                             
\end{center}

\begin{center}
\begin{deluxetable}{cccccccc}
\tablewidth{0pt}
\tablecaption{Phase Shifts and Flux Scale factors\label{tab-shifts}}
\tablecolumns{8}
\tablehead{
\colhead{Quantity}                       &
\colhead{RR Cet}                         &
\colhead{RV Oct}                         &
\colhead{SS Leo}                         &
\colhead{ST Leo}                         &
\colhead{W Tuc}                          &
\colhead{WY Ant}                         &
\colhead{X Ari}                           
}
\startdata
\multicolumn{8}{c}{flux scale factors} \\
H$\alpha$                 & 1.136 & 1.085 & 0.910 & 1.450 & 1.104 & 1.040 & 0.770 \\
\species{He}{i} 5876~\AA\ & 1.135 & 0.850 & 0.780 & 1.200 & 1.640 & 1.020 & 1.000 \\
\multicolumn{8}{c}{phase shifts $\Delta\phi$} \\
H$\alpha$                 &    0.0060 & $-$0.0095 & $-$0.0035 & $-$0.0220 & $-$0.0035 &    0.0090 &    0.0061 \\
\species{He}{i} 5876~\AA\ &    0.0080 & $-$0.0050 & $-$0.0005 & $-$0.0180 &    0.0005 &    0.0085 &    0.0053 \\
\enddata

\end{deluxetable}
\end{center}

\begin{center}
\begin{deluxetable}{lrrrrrr}
\tablewidth{0pt}
\tablecaption{Kinematic Properties of RRab Stars During Primary Light Rise
     \tablenotemark{a}\label{tab-kinrrab}}
\tablecolumns{7}
\tablehead{
\colhead{Star}                           &
\colhead{dR/Dt}                          &
\colhead{dR/dt}                          &
\colhead{dR/dt}                          &
\colhead{\sigmax}                        &
\colhead{\sigmax}                        &
\colhead{\sigmax}                         
}
\startdata
       & H$\alpha$(B) & H$\alpha$(R) & H$\alpha$(B-R) & H$\alpha$(B) & H$\alpha$(R) & H$\alpha$(B-R) \\
RR Cet      &  86.4 & $-$71.6 & 158.0 & 85.8 & 57.8 & 28.0 \\
RV Oct      &  93.6 & $-$72.1 & 165.7 & 81.7 & 57.8 & 23.9 \\
SS Leo      &  87.5 & $-$76.2 & 163.7 & 83.2 & 59.1 & 24.1 \\
W Tuc       & 109.9 & $-$71.9 & 181.8 & 94.0 & 72.3 & 21.7 \\
WY Ant      &  87.9 & $-$73.3 & 161.2 & 87.3 & 64.8 & 22.5 \\
X Ari       &  88.4 & $-$74.2 & 162.6 & 87.8 & 60.1 & 27.7 \\
mean        &  92.3 & $-$73.2 & 165.5 & 86.6 & 62.0 & 24.7 \\
$\sigma$    &   8.2 &     1.6 &   7.7 &  3.9 &  5.2 &  2.4 \\
            &       &         &       &      &      &      \\
            & \species{He}{i} & \species{He}{i} & \species{He}{i} & \species{He}{i} & \species{He}{i} & \species{He}{i} \\
RR Cet      &  57.8 & $-$87.7 & 145.5 & 43.8 & 28.6 & 15.2 \\
RV Oct      &  58.0 & $-$73.8 & 131.8 & 39.8 & 23.3 & 16.5 \\
SS Leo      &  53.0 & $-$73.5 & 126.5 & 40.6 & 26.5 & 14.1 \\
W Tuc       &  52.2 & $-$69.6 & 121.8 & 45.2 & 36.1 &  9.1 \\
WY Ant      &  48.2 & $-$74.6 & 122.8 & 39.6 & 39.3 &  0.3 \\
X Ari       &  53.1 & $-$62.9 & 116.0 & 51.6 & 43.3 &  8.3 \\
mean        &  53.7 & $-$73.7 & 127.4 & 43.4 & 32.9 & 10.6 \\
$\sigma$    &   3.4 &     7.4 &   9.4 &  4.2 &  7.2 &  5.5 \\
\enddata

\tablenotetext{a}{Quantities measured at phase $\phi$~$\simeq$~0.97}
                  
\end{deluxetable}
\end{center}

\begin{center}
\begin{deluxetable}{lrrrrrrrrr}
\tablewidth{0pt}
\tablecaption{Regression constants for dR/dt and \sigmax\label{tab-regress}}
\tablecolumns{10}
\tablehead{
\colhead{Star}                           &
\colhead{feature}                        &
\colhead{dR/Dt(B)}                       &
\colhead{dR/dt(B)}                       &
\colhead{dR/dt(R)}                       &
\colhead{dR/dt(R)}                       &
\colhead{\sigmax(B)}                     &
\colhead{\sigmax(B)}                     &
\colhead{\sigmax(R)}                     &
\colhead{\sigmax(R)}                     \\
\colhead{}                               &
\colhead{}                               &
\colhead{a0}                             &
\colhead{a1}                             &
\colhead{a0}                             &
\colhead{a1}                             &
\colhead{a0}                             &
\colhead{a1}                             &
\colhead{a0}                             &
\colhead{a1}                            
}
\startdata
RR Cet & H$\alpha$       &     186.554 & $-$103.250 & 523.387 & $-$613.388 & $-$204.589 &   299.338 &    258.439 & $-$206.717 \\
RV Oct & H$\alpha$       &   $-$34.041 &    131.576 & 249.885 & $-$331.951 &  $-$72.621 &   159.074 &    201.159 & $-$147.791 \\
SS Leo & H$\alpha$       &  $-$216.277 &    313.160 & 614.894 & $-$709.663 & $-$249.830 &   343.312 &    272.169 & $-$219.620 \\
W Tuc  & H$\alpha$       &      12.285 &    100.665 & 134.548 & $-$212.862 &     56.774 &    38.419 &     93.955 &  $-$22.313 \\
WY Ant & H$\alpha$       &      59.292 &     31.410 & 283.985 & $-$368.308 &  $-$76.043 &   168.379 &    258.183 & $-$199.415 \\
X Ari  & H$\alpha$       &  $-$146.453 &    242.098 & 137.306 & $-$218.010 &     35.781 &    53.628 &    326.617 & $-$274.785 \\
RR Cet & \species{He}{i} &  $-$850.430 &    936.342 & 572.532 & $-$680.664 &  $-$78.794 &   126.436 & $-$217.945 &    254.132 \\
RV Oct & \species{He}{i} & $-$1020.236 &   1111.608 & 313.749 & $-$399.558 &     84.620 & $-$46.234 &  $-$32.516 &     57.567 \\
SS Leo & \species{He}{i} &  $-$865.432 &    946.876 & 614.894 & $-$709.663 & $-$124.863 &   170.547 &   $-$6.733 &     35.940 \\
W Tuc  & \species{He}{i} &  $-$809.648 &    888.546 & 360.180 & $-$443.044 & $-$148.486 &   199.692 &  $-$96.769 &    136.933 \\
WY Ant & \species{He}{i} &  $-$689.407 &    760.464 & 464.721 & $-$556.043 & $-$107.779 &   151.970 & $-$202.714 &    249.459 \\
X Ari  & \species{He}{i} &  $-$781.674 &    860.612 & 437.561 & $-$515.920 & \tablenotemark{a} & \tablenotemark{a} & \tablenotemark{b} & \tablenotemark{b} \\
\enddata
\tablenotetext{a}{Quadratic fit: \sigmax\ = $-$8531 + 17733$\phi$ $-$9160$\phi^2$}
\tablenotetext{b}{Quadratic fit: \sigmax\ = $-$10910 + 22426$\phi$ $-$11478$\phi^2$}
\end{deluxetable}
\end{center}

\end{document}